\documentclass[twocolumn,preprintnumbers,amsmath,amssymb,superscriptaddress,aps]{revtex4}

\usepackage{graphicx}
\usepackage{bm}
\usepackage{SIunits}
\usepackage{ulem}
\usepackage{color}


\begin{document}

\title{Theory of harmonic dissipation in disordered solids}

\author{T. Damart}
\affiliation{Institut Lumi\`ere Mati\`ere, UMR 5306 Universit\'e Lyon 1-CNRS, Universit\'e de Lyon, F-69622 Villeurbanne, France}
\author{A. Tanguy}
\affiliation{Laboratoire de M\'ecanique des Contacts et des Structures, UMR 5259 Institut National des Sciences Appliqu\'ees de Lyon-CNRS, F-69621 Villeurbanne, France}
\author{D. Rodney}
\affiliation{Institut Lumi\`ere Mati\`ere, UMR 5306 Universit\'e Lyon 1-CNRS, Universit\'e de Lyon, F-69622 Villeurbanne, France}

\begin{abstract}
Mechanical spectroscopy, i.e. cyclic deformations at varying frequencies, is used theoretically and numerically to measure dissipation in model glasses. From a normal mode analysis, we show that in the high-frequency THz regime where dissipation is harmonic, the quality factor (or loss angle) can be expressed analytically. This expression is validated through non-equilibrium molecular dynamics simulations applied to a model of amorphous silica (SiO$_2$). Dissipation is shown to arise from non-affine relaxations triggered by the applied strain through the excitation of vibrational eigenmodes that act as damped harmonic oscillators. We discuss an asymmetry vector field, which encodes the information about the structural origin of dissipation measured by mechanical spectroscopy. In the particular case of silica, we find that the motion of oxygen atoms, which induce a deformation of Si-O-Si bonds is the main contributor to harmonic energy dissipation.
\end{abstract}

\maketitle

\section{INTRODUCTION}

Mechanical losses through energy dissipation is a limiting factor central to the design of high-precision devices, such as micro- and nanoelectromechanical systems (MEMS/NEMS)~\cite{lifshitz2000,houston2002,
li2007} and even highly sensitive interferometers such as gravitational wave detectors, whose resolution is currently controlled by the energy dissipated in the oxide glass coatings of the mirrors~\cite{saulson1990,flaminio2010,granata2013}. While the source of energy dissipation in crystals can be traced back to crystalline defects~\cite{seeger1957}, dissipation in glasses can take many forms and may involve diverse phenomena such as thermally-activated relaxations~\cite{jackle1972,phillips1987,vacher2005, hamdan2014}, Zener thermoelastic damping~\cite{zener2, zener3}, Akhiezer damping~\cite{fabian1999,kunal2011,carini2014}, Rayleigh scattering~\cite{ruocco1999,ganter2010,liang2016,baldi2016} and more, depending on the frequency range of interest.  

At high frequencies (THz), dissipation in glasses arises from the attenuation of collective vibrational excitations~\cite{vacher2005, masciovecchio2006,baldi2014}.
The latter are most often studied through the dynamical structure factor, $S(q,\omega)$, the space- and time-Fourier transform of the density-density correlation function~\cite{hansen2006}, which is measured both experimentally using inelastic x-ray scattering (IXS) (see for instance Refs.~\onlinecite{foret1996,ruocco2001, masciovecchio2006, baldi2014, benassi1996, devos2008, baldi2016}), and numerically with molecular dynamics (MD) (see for instance Refs.~\onlinecite{grest1982, taraskin1997, dellanna1998,ruocco2000, shintani2008, monaco2009, marruzzo2013, beltukov2013, beltukov2016, liang2016}). 

At small wave vectors, typically below 1-3 nm$^{-1}$ in amorphous silica~\cite{horbach2001, baldi2010, liang2016}, the spectrum of $S(q,\omega)$ shows a peak at a well-defined frequency, an evidence that in this regime, glasses support propagating vibrational modes, similar to crystalline phonons, but with a damping related to the finite width of the excitation peak. The latter is therefore naturally fitted as a damped harmonic oscillator (DHO) to extract the excitation frequency $\Omega(q)$ and attenuation (or linewidth), $\Gamma(q)$. Dissipation, as measured from the quality factor associated with the loss angle, is then obtained as $Q^{-1}=\Gamma/\Omega$, a relation which holds at low damping for instance in Zener's standard linear solid~\cite{parke1966}. 

However, for wave vectors larger than a few nm$^{-1}$, in the region leading to the boson peak (BP)~\cite{elliott1984, buchenau1986, binder2005, duval2007}, glasses exhibit a strong damping with an attenuation increasing rapidly with frequency. 
$\Gamma \propto \Omega^\alpha$, with $\alpha \sim 4$ in 3D, in both experiments~\cite{foret1996, ruffle2006, baldi2010, ferrante2013} and MD simulations~\cite{schirmacher1998, grigera2001, monaco2009, marruzzo2013, degiuli2014, liang2016} (a logarithmic correction was identified very recently in Ref.~\onlinecite{gelin2016}). In this regime, damping is mainly of harmonic origin and is controlled by the structural disorder in the glass \cite{dellanna1998,ruocco1999, ruocco2000, degiuli2014}. 
More specifically, the acoustic vibrations undergo a Rayleigh type of scattering by the elastic heterogeneities in the glass that are correlated on the same nanometer scale as the wavelength of the acoustic vibrations at the boson peak~\cite{tanguy2002, tsamados2009, ganter2010, wagner2011, tanguy2015, marruzzo2013}. As a result, the phonon mean free path ($\propto 1/\Gamma$) decreases rapidly with frequency and becomes comparable to the wavelength, thus reaching the Ioffe-Regel (IR) limit ($\Gamma = \Omega/\pi$)~\cite{ioffe1960}, above which the notion of phonon with a well-defined wave vector is inapplicable~\cite{allen1999, beltukov2013}.

Above the IR limit, $S(q,\omega)$ shows a very broad peak, which results from the convolution of several excitations that cannot technically, nor should theoretically, be fitted as a DHO, as recently mentioned in the conclusions of Refs.~\onlinecite{buchenau2014, baldi2016, liang2016}. How can we then describe dissipation in glasses above the Ioffe-Regel limit? This question is addressed theoretically in the present paper by measuring dissipation directly using mechanical spectroscopy. With this technique, the loss (or internal friction) angle $\phi$ is measured between an imposed sinusoidal strain and the resulting internal stress, yielding the energy dissipation $Q^{-1}=\tan \phi$. This approach is widely used experimentally in the Hz to kHz regime to study glasses~\cite{qiao2013, liu2015}, liquids and soft matter systems~\cite{rubinstein2003, coussot2005}, but has also been used numerically at higher frequencies in MD simulations~\cite{vladkov2006, tseng2010, wittmer2015}. Here we show that in the high-frequency regime of harmonic dissipation, the quality factor $Q^{-1}$ can be expressed analytically, allowing to analyze in details the features that control dissipation in a glass, both below and above the IR limit.

In the present work, we study dissipation using a combination of non-equilibrium MD simulations applied to a model of amorphous silica (SiO$_2$), and an analytical expression obtained in the harmonic approximation. Details of the simulations and the analytical calculations are given in Sec.~\ref{methodo}. The results of the simulations are compared to experimental data and the analytical expression in Sec.~\ref{results}. Finally, in Sec.~\ref{properties}, we discuss the properties of dissipation deduced from the analytical expression.

\section{METHODOLOGY}
\label{methodo}

\subsection{Glass model}

\begin{figure}
 \includegraphics [width=0.5\textwidth]{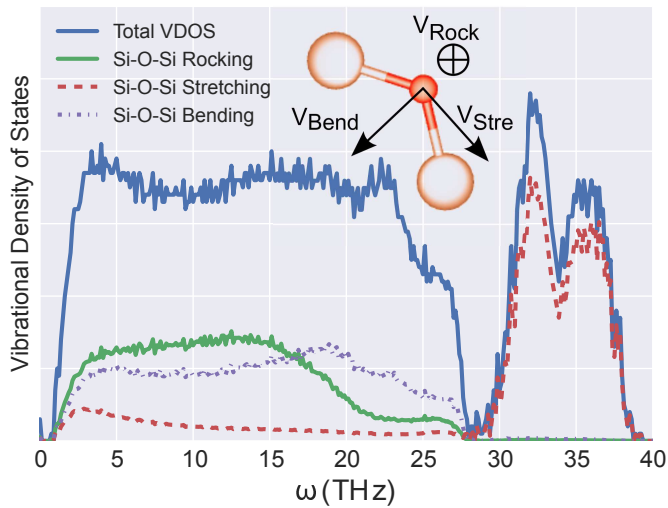}
  \caption{Vibrational density of states of the SiO$_2$ model, decomposed on the rocking, stretching and bending motions of the Si-O-Si bonds. The inset shows a sketch of a Si-O-Si bond with the vectors used to decompose the bond deformation into bending, stretching and rocking components.}
  \protect\label{dos}
\end{figure}

The glass model considered for this study is amorphous silica (SiO$_2$), whose structure and properties are well known\cite{bruckner1970, zeller1971, vollmayr1996, taraskin1997, taraskin1999, rahmani2003,koziatek2015}. The MD simulations were performed in 3D cubic cells, with a side $L=34.77~\angstrom$, containing 1,000 Si atoms and 2,000 O atoms (the corresponding density is $2.4$~g/cm$^3$). Glassy structures were obtained by quenching a liquid melt by MD at a constant quench rate of $10^{10}$~K.s$^{-1}$. Interactions between particles are described using a standard van Beest, Kramer, and van Santen (BKS) pair potential~\cite{beest1990}, with the long-range Coulomb interactions screened using a Wolf truncation. We used the cut-off function proposed by Carr\'{e}~\textit{et al.}~\cite{carre2007} and already employed in Refs.~\onlinecite{mantisi2012, koziatek2015} to model SiO$_2$ glasses. Periodic boundary conditions are applied in all directions. An example of configuration is shown in the inset of Fig.~\ref{Cyc}. 
The corresponding vibrational density of states (VDOS) is represented in Fig.~\ref{dos} along with the partial VDOS for oxygen atoms projected on the rocking, stretching and bending motions of the Si-O-Si bonds, as done in Refs.~\onlinecite{taraskin1997b, giustino2006, shcheblanov2015}.

\subsection{Mechanical spectroscopy}

Mechanical spectroscopy was simulated by imposing a cyclic deformation to the simulation cell. In the following, we mainly consider the case of isostatic deformations. We also performed simple shear deformations, but they resulted in qualitatively similar results and will not be presented here.

For isostatic deformations, the simulation cell is subjected in the $X$, $Y$ and $Z$ directions to a sinusoidal applied strain $\epsilon(t) = \epsilon_0 \sin(\omega t)$ with a frequency $\omega/2\pi$ varying from 0.1 to 50 THz and an amplitude $\epsilon_0=0.007$ chosen such that the deformation remains elastic in the quasistatic limit. The system is thermostated in order to dissipate the heat produced during the deformation cycles and maintain a constant temperature, which was varied from 10~K to 700~K. We compared different thermostats (Andersen, Nose-Hoover, Langevin)~\cite{allen1987} with different strengths but did not find any marked influence. In the following, we will consider a Langevin thermostat, which allows for analytical calculations developed in Sec. \ref{subsec:analytic}. Atomic trajectories are integrated using SLLOD equations for isostatic tractions and compressions~\cite{allen1987}:
\begin{eqnarray}
	\dot{r}_i^{\alpha} &=& \frac{p_i^{\alpha}}{m_i} + \dot{\epsilon} r_i^{\alpha} \nonumber \\
	\dot{p}_i^{\alpha} &=& F_i^{\alpha} - \dot{\epsilon} p_i^{\alpha} - \gamma p_i^{\alpha} +F_{th},
	\label{sllod0}
\end{eqnarray}
where $r_i^{\alpha}$ is the current position of atom $i$ in direction ${\alpha}$ in the deformed simulation cell, $p_i^{\alpha}$ its momentum and $F_i^{\alpha}$ the force coming from the interatomic potential. The Langevin friction, $\gamma$, is related to the random force, $F_{th}$, through the fluctuation-dissipation theorem. $\gamma$ was varied between 0.1 and 10 THz. Below 0.1 THz, the thermostat is too weak to maintain a constant temperature and above 10 THz, the forcing is too strong and affected the dynamics of the glass (the influence of the friction parameter will be further discussed in Sec.~\ref{sec:discuss}). The time step of the simulations was 1~fs when the forcing frequency was $1$~THz or below. Above 1~THz, the time step was set to $10^{-3}/\omega$ in order to maintain a constant strain increment per simulation step. 

\begin{figure}
 \includegraphics [width=0.5\textwidth]{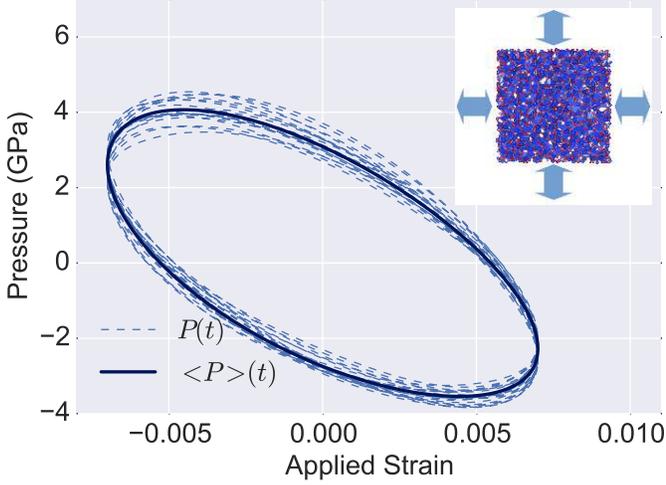}
  \caption{Evolution of the hydrostatic pressure during cycles of isostatic applied strain. The dashed curves show the instantaneous pressure. The solid curve is the average computed over 250 cycles.}
  \protect\label{Cyc}
\end{figure}

We follow simultaneously the time-evolution of the pressure $P(t)$, which, in the stationary regime, is a periodic function of same period as the applied strain ($T = 2 \pi/\omega$) with cycle-dependent fluctuations illustrated in Fig.~\ref{Cyc}. In the following, we consider the smooth periodic part of the pressure averaged over multiple cycles:

\begin{equation}
	\langle P \rangle(t) = \lim\limits_{N \rightarrow +\infty} \frac{1}{N} \sum_{n=0}^N P(t+nT).
\end{equation} 

The dissipation $Q^{-1}$ is related to the loss angle $\phi$ between the pressure $\langle P \rangle(t)$ and the applied strain $\epsilon(t)$:

\begin{eqnarray}
	Q^{-1}(\omega) &=& \tan (\phi) \nonumber \\
	&=& \frac{1}{\omega}\frac{\int_0^T \langle P \rangle(t) \dot{\epsilon} (t) \, \mathrm{d}t}{\int_0^T \langle P \rangle(t) \epsilon (t) \, \mathrm{d}t}
    \label{eq:dissip}
\end{eqnarray}

The spectroscopic simulations were performed during 300 deformation cycles, the value of the energy dissipation usually converging after about 50 cycles, after which the glass enters a stationary regime where all measurements were carried out. 

\subsection{Harmonic approximation}
\label{subsec:harmo}

In order to separate harmonic and anharmonic effects, mechanical spectroscopy was also applied to harmonic systems, where the interaction between particles was described using the dynamical matrix of the system, $\tilde{D}_{ij}^{\alpha \beta} = \frac{1}{\sqrt{m_im_j}}\frac{\partial^2E}{\partial r_i^{\alpha}\partial r_j^{\beta}} = \frac{1}{\sqrt{m_im_j}}D_{ij}^{\alpha \beta}$, where indices $i$ and $j$ refer to atoms and ${\alpha}$ and ${\beta}$ to Cartesian coordinates. The potential energy, $E$, is computed using the Wolf-truncated BKS potential, which is an analytical pair potential and thus yields an analytical expression for the dynamical matrix. The potential energy of the system was then approximated as:
\begin{equation}
	E=-\frac{1}{4}\sum_{i\alpha j\beta}D_{ij}^{\alpha \beta}(r_{ij}^{\beta}-R_{ij}^{\beta})(r_{ij}^{\alpha}-R_{ij}^{\alpha})
	\label{energy}
\end{equation}
where $r_{ij}^\alpha$ (resp. $R_{ij}^\alpha$) is the separation between atoms $i$ and $j$ in direction ${\alpha}$ in the deformed  (resp. initial) cell, using the minimum image convention to account for the periodic boundary conditions. The corresponding expressions for the atomic forces and pressure are given in Appendix~\ref{sec:analytic}.

\subsection{Analytic expression}
\label{subsec:analytic}

The dissipation measured numerically in Eq.~\ref{eq:dissip} can be expressed analytically in the harmonic approximation and linear response regime when a Langevin thermostat is assumed. The calculations are detailed in Appendix~\ref{sec:analytic} in the case of isostatic deformations. Generalization to arbitrary deformations (for instance shear) is straightforward.

Dissipation is calculated as the ratio of the imaginary and real parts of the complex modulus, which relates the Fourier transforms of the periodic applied strain $\epsilon_{\kappa\xi}$ to the cycled-averaged internal stress $\langle \sigma_{\alpha \beta} \rangle$. For that, the cycled-averaged stress in the harmonic approximation is projected on the normal modes of the glass and expressed as the sum of an affine and non-affine contribution:
\begin{equation}
\langle \sigma_{\alpha \beta} \rangle(\omega) =  C_{\alpha\beta\kappa\xi}^\infty \epsilon_{\kappa\xi}(\omega)-\frac{2}{V_0} \sum_m C_m^{\alpha\beta} \langle s_m \rangle(\omega),
\label{stress0}
\end{equation}
where $\omega/2\pi$ is the forcing frequency, $s_m$ the mass-scaled amplitude of the $m^{th}$ normal mode and $V_0$ the volume of the reference undeformed cell. The first term in the RHS of Eq. \ref{stress0} is the affine Born contribution, with $C_{\alpha\beta\kappa\xi}^\infty$ the affine elastic modulus obtained when all atoms are forced to follow the macroscopic applied uniform deformation $\epsilon_{\kappa\xi}$:
\begin{equation}
C_{\alpha\beta\kappa\xi}^\infty = -\frac{1}{2V_0} \sum_{ij}  \Big[ D_{ij}^{\alpha\kappa} R_{ij}^{\beta} + D_{ij}^{\beta\kappa} R_{ij}^{\alpha} \Big]  R_{ij}^{\xi}.
\end{equation}

The second term in the RHS of Eq. \ref{stress0} is the non-affine contribution, which is expressed as a sum over the normal modes of the system:
\begin{equation}
C_m^{\alpha\beta} = \frac{1}{2} \sum_{ij\kappa}  \Big[ D_{ij}^{\alpha \kappa} R_{ij}^{\beta}+D_{ij}^{\beta \kappa} R_{ij}^{\alpha} \Big]  \frac{e_j^{\kappa}(m)}{\sqrt{m_j}}.
\label{Cmgeneral}
\end{equation}

Here, $e_j^\kappa(m)$ is the component on atom $j$ and direction $\kappa$ of the $m^{th}$ eigenvector of the mass-scaled dynamical matrix $\tilde{D}$, with corresponding eigenfrequency $\omega_m$. 

The temporal Fourier transform of the non-affine displacement, $\langle s_m\rangle (\omega)$, is expressed by projecting the linearized SLLOD equations of motion in Eq. \ref{sllod0} on the normal modes of the system (see Appendix~\ref{sec:analytic} for details):
\begin{equation}
\langle s_m\rangle (\omega) = \frac{C_m^{\kappa\xi}}{\omega_m^2-\omega^2+i\gamma\omega} \epsilon_{\kappa\xi}(\omega)
\label{decompo}
\end{equation}

From Eq. \ref{stress0}, the resulting complex modulus is thus:
\begin{equation}
C_{\alpha\beta\kappa\xi}(\omega) = C_{\alpha\beta\kappa\xi}^\infty -\frac{2}{V_0} \sum_m \frac{C_m^{\alpha\beta}C_m^{\kappa\xi}}{\omega_m^2-\omega^2+i\gamma\omega},
\label{elastmod}
\end{equation}
which involves the response function of the normal modes, $1/(\omega_m^2-\omega^2+i\gamma\omega)$, broadened by the Langevin thermostat through the $i\gamma\omega$ term. Building on the decomposition of the elastic constants into affine and non-affine contributions first proposed by Lutsko~\cite{lutsko1989}, Lema\^itre and Maloney obtained an expression similar to Eq.~\ref{elastmod} to analyze the visco-elastic response of disordered solids~\cite{lemaitre2006}. The static limit of this expression ($\omega =0$) was used by these authors and Zaccone {\it et al.}~\cite{zaccone2011} to study the effect of non-affine relaxations on the elasticity of glasses.

In case of isostatic deformations of main interest here, the above equations adopt a more compact form (see Appendix~\ref{sec:analytic} for details), with the complex bulk modulus relating the Fourier transform of the average pressure $\langle P \rangle (\omega)$ to the applied strain $\epsilon(\omega)$:
\begin{equation}
	K(\omega) = K^\infty - \frac{2}{V_0} \sum_m \frac{C^2_m}{\omega_m^2-\omega^2+i\gamma\omega}.
\end{equation}
The affine bulk modulus and mode-dependent non-affine term are respectively:
\begin{equation}
\begin{aligned}
	K^\infty &= - \frac{1}{V_0}\sum_{i\alpha j\beta}D_{ij}^{\alpha \beta}R_{ij}^{\beta} R_{ij}^{\alpha}, \\
	C_m &= \sum_{i\alpha j\beta}D_{ij}^{\alpha \beta}R_{ij}^{\alpha} \frac{e_j^{\beta}(m)}{\sqrt{m_j}}.
	\label{Cm}
    \end{aligned}
\end{equation}
Finally, we obtain the expression of the dissipation produced by isotropic deformations:
\begin{equation}
	Q^{-1}(\omega) =\frac{\sum_m C^2_m \frac{\omega\gamma}{(\omega_m^2-\omega^2)^2+(\gamma\omega)^2}}{\frac{V_0}{2}K^\infty - \sum_m C^2_m \frac{\omega_m^2-\omega^2}{(\omega_m^2-\omega^2)^2+(\gamma\omega)^2}}
    \label{analytic}
\end{equation}

In the next Section, we compare this analytic expression of the dissipation with numerical calculations and discuss the physical insights gained from this expression on the origin of dissipation in glasses below and above the Ioffe-Regel limit. 

\section{\label{sec:discuss}SIMULATION RESULTS}
\label{results}

\begin{figure}[h]
 \includegraphics [width=0.5\textwidth]{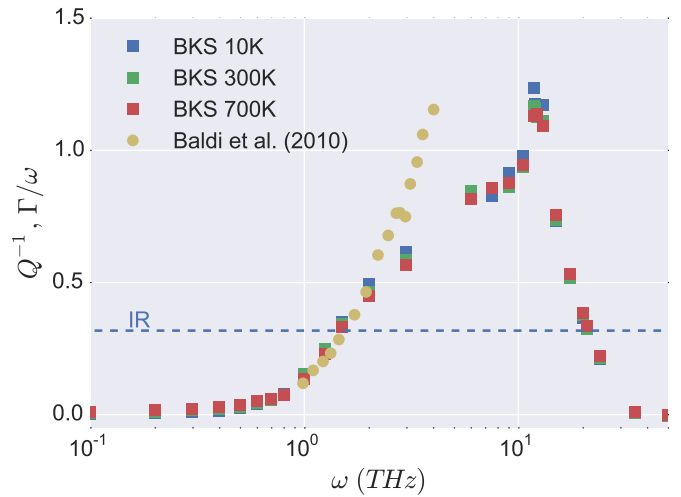}
  \caption{Energy dissipation as a function of frequency in amorphous SiO$_2$ modeled with the full non-linear BKS potential at three different temperatures. The friction of the Langevin thermostat was set to 1 THz. The yellow circles are experimental data obtained by fitting the excitation peak of x-ray scattering spectra with the DHO model~\cite{baldi2010}.}
  \protect\label{Q1}
\end{figure}

\subsection{Full non-linear calculations}

The energy dissipation obtained by mechanical spectroscopy in the present amorphous SiO$_2$ model is presented in Fig.~\ref{Q1} at three different temperatures, for a range of frequencies, which broadly covers that of the vibrational density of states of the glass in Fig.~\ref{dos}.  Dissipation is numerically zero below 0.1~THz, increases up to about 1.2 at 12~THz and decreases back to zero above 27~THz. The three sets of data obtained at 10, 300 and 700~K are superimposed. This temperature independence is a strong indication that, as expected from previous works~\cite{dellanna1998, taraskin1999, ruocco2000, taraskin2000}, dissipation is harmonic in the present range of high frequencies.

In Fig.~\ref{Q1}, are also plotted the experimental attenuation data $\Gamma/\omega$ of Baldi \textit{et al.}\cite{baldi2010}, obtained using a DHO fit of the dynamical structure factor of vitreous SiO$_2$. The very good agreement between the experimental and numerical data below the Ioffe-Regel (IR) limit, $\Gamma/\omega=1/\pi$, confirms the strong connection between the quality $Q^{-1}$ and attenuation $\Gamma/\omega$ factors. However, we note that this link is difficult to justify theoretically, even in the harmonic approximation considered below~\cite{parke1966}.

We see in Fig.~\ref{Q1} that above the IR limit, the experimental data overestimate the numerical dissipation. This discrepancy might be expected for two reasons. First, simple models like Zener's standard linear solid~\cite{parke1966}, predict that the dissipation and attenuation factors match only in the limit of low dissipation, while at large dissipation, the attenuation factor overestimates the quality factor. Second, above the IR limit, the dynamical structure factor contains several excitation peaks that cannot be fitted by a simple DHO model~\cite{christie2007, baldi2014, damart2015}.

\begin{figure}
 \includegraphics [width=0.5\textwidth]{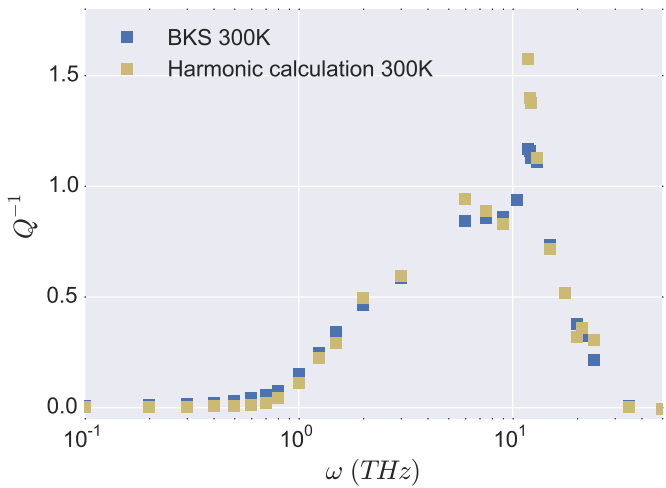}
  \caption{Energy dissipation as a function of frequency computed with the non-linear BKS potential and the harmonic approximation of Eq.~\ref{energy}. The same thermostat friction of 1 THz was used in all calculations.}
  \protect\label{Q2}
\end{figure}

\subsection{Harmonic approximation}

To confirm the harmonic origin of energy dissipation in the present range of frequencies, we applied mechanical spectroscopy to the same sample but with the interactions between particles described using the dynamical matrix of the equilibrium configuration, as explained in Sec.~\ref{subsec:harmo}. The resulting energy dissipation is compared in Fig.~\ref{Q2} with the full non-linear BKS calculations. The very good agreement between both calculations confirms the harmonic origin of dissipation in this frequency range. We note that at low frequencies, typically below 1 THz, the harmonic calculations find a dissipation systematically lower than the non-linear model, an indication that anharmonic effects may play a role in this region. However, we will see below that dissipation measurements are strongly affected by the thermostat in this low-frequency regime.

\begin{figure}
 \includegraphics [width=0.5\textwidth]{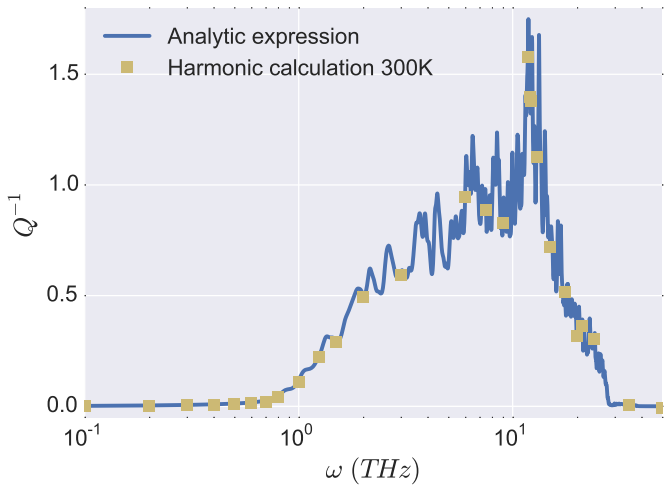}
  \caption{Energy dissipation as a function of frequency computed with the harmonic approximation compared to the analytical expression of Eq.~\ref{analytic}. The same thermostat friction of 1 THz is used in both cases.}
  \protect\label{Q3}
\end{figure}

Finally, we compare in Fig.~\ref{Q3} the harmonic simulations to the analytical expression in Eq.~\ref{analytic}. The perfect agreement between both approaches, even in the regions where the dissipation fluctuates rapidly (e.g. near 10 THz), confirms the validity of the analytic calculations of Sec. \ref{subsec:analytic}. The expression of energy dissipation in Eq.~\ref{analytic} also shows directly that dissipation in the harmonic regime is independent of the temperature and the strain amplitude ($\epsilon_0$). $Q^{-1}$ however depends on the friction parameter $\gamma$ of the Langevin thermostat, a point detailed in the following Section, where we also address other properties of dissipation deduced from the analytical expression. 

\section{PROPERTIES OF HARMONIC DISSIPATION}
\label{properties}

\subsection{Physical interpretation}

Focusing on the numerator of Eq.~\ref{analytic}, which mostly controls the shape of the dissipation spectrum, we see that $Q^{-1}$ is expressed as a sum of contributions coming from the vibrational eigenmodes. Each contribution is the product of the square of the non-affine coefficient $C_m$ (Eq.~\ref{Cm}) with a Lorentzian centered on the mode eigenfrequency $\omega_m$, with a width fixed by the Langevin friction $\gamma$.

Physically, dissipation arises because the deformation applied to the cell triggers non-affine relaxations (Eq.~\ref{stress0}) that are supported by the eigenmodes of the system (Eq.~\ref{decompo}). Since the latter are harmonic oscillators damped by the thermostat, they induce a lag in the non-affine stress contribution, which is maximum when the forcing frequency equals the mode eigenfrequency. The coupling coefficient, $C_m^{\alpha\beta}$, reflects the sensitivity of the stress on the amplitude of mode $m$, since from Eq.~\ref{stress0}, $C_m^{\alpha\beta} \propto \partial \sigma_{\alpha \beta} / \partial \langle s_m \rangle$. The modes that dissipate most are therefore those that produce large non-affine stress relaxations and resonate with the forcing frequency.


\subsection{Influence of the Langevin friction parameter}

\begin{figure}
 \includegraphics [width=0.5\textwidth]{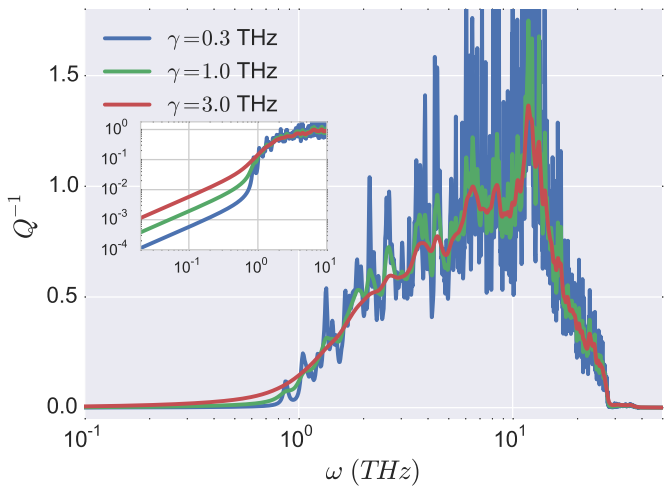}
  \caption{Energy dissipation as a function of frequency computed using the analytical expression for three different thermostat frictions: 0.3, 1 and 3~THZ. The inset shows a log-log view of the low frequency region.}
  \protect\label{Q4}
\end{figure}

The dissipation in Eq.~\ref{analytic} depends on the friction parameter $\gamma$ of the Langevin thermostat, which might appear as an artifact since $\gamma$ is a numerical parameter with no physically defined value. However, we argue below that except in the region $\omega<\gamma$, the shape and main features of the dissipation spectrum do not depend on $\gamma$.

The effect of a finite value of $\gamma$ is to broaden the response function of the eigenmode oscillators (Eq.~\ref{decompo}). As a result, the numerator of the dissipation in Eq.~\ref{analytic} is expressed as a weighted average of the non-affine coefficients $C_m$ over a frequency window of order $\gamma$. As will be emphasized in Fig.~\ref{C2}, $C_m$ varies rapidly from mode to mode. Therefore, when $\gamma$ is small, the non-affine parameter $C_m$ is not averaged over a large enough window and the dissipation shows rapid fluctuations, as seen in Fig.~\ref{Q3}. However, when $\gamma$ increases and $C_m$ is averaged over more modes, the dissipation spectrum becomes smoother but retains the same shape and features, as shown in Fig.~\ref{Q4}, even near the peak of dissipation at 10 THz. This is typically true as long as $\omega$ remains in the frequency spectrum of the density of states and $\omega > \gamma$. Indeed, in the limit $ \omega < \min(\gamma,\omega_m)$, Eq.~\ref{analytic} predicts $Q^{-1} \propto \gamma \omega$. This is visible in the inset of Fig.~\ref{Q4}, where the dissipation below typically 1 THz scales with the the frequency and friction. In this region outside the VDOS, the slow decay when $\omega \to 0$ is an artifact of the finite width of the Lorentzian and therefore, of the finite-friction thermostat. 
Equivalently, we can say that the fluctuations seen in Fig.~\ref{Q3} are a finite size effect, due to the fact that in the small systems considered here, there are not enough modes to obtain a smooth average of $C_m$. Larger systems with denser eigenfrequency spectra would show smoother dissipations at fixed $\gamma$. However, considering larger systems is difficult since diagonalizing  the dynamical matrix becomes rapidly very computationally intensive. 

\begin{figure}
 \includegraphics [width=0.5\textwidth]{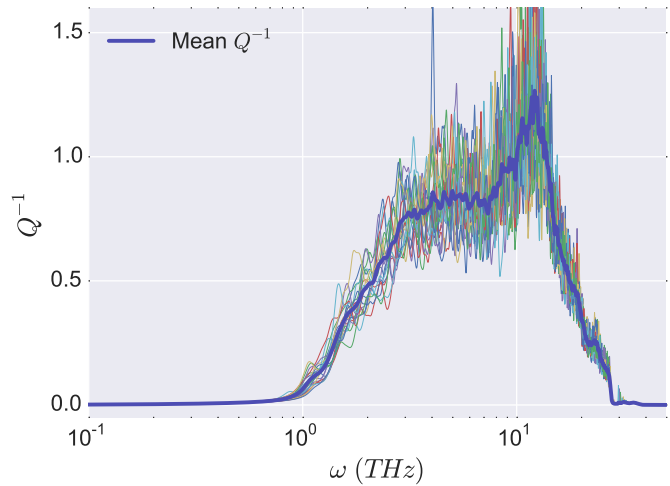}
  \caption{Energy dissipation as a function of frequency calculated using the analytic expression for 20 different SiO$_2$ glasses with a thermostat frequency $\gamma=1$~THz.}
  \protect\label{Q_MULTI}
\end{figure}

Another way to limit the fluctuations is to average the dissipation spectrum over independent SiO$_2$ glassy configurations of same size, as done in Fig.~\ref{Q_MULTI}. Fluctuations between different configurations are obvious, but the general shape remains the same and the average curve shows the same features as seen in Figs.~\ref{Q1} and~\ref{Q4}.

\subsection{Properties of $C_m$}

\begin{figure}
 \includegraphics [width=0.5\textwidth]{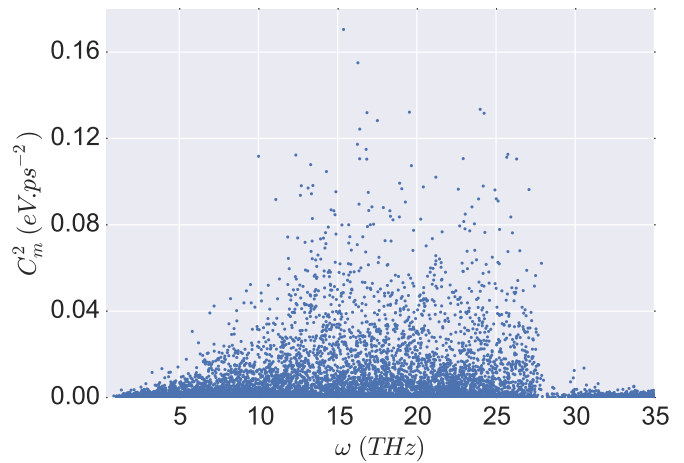}
  \caption{Square of the mode-dependent non-affine coefficient, $C_m^2$, computed from Eq.~\ref{Cm}, as a function of the modes eigenfrequency.}
  \protect\label{C2}
\end{figure}

From Eqs.~\ref{stress0} and \ref{decompo}, every eigenmode $m$ has an influence on the total stress, through a non-affine contribution (departure of the mode from the affine macroscopic imposed displacement) whose amplitude is fixed by the mode-dependent parameter $C_m^{\alpha \beta}$. Fig.~\ref{C2} shows $C_m^2 = (\sum_\alpha C_m^{\alpha\alpha}/3)^2$, the square of the coupling parameter for isostatic deformations, as a function of the mode frequency. We see that (1) the coupling parameter varies rapidly from mode to mode, (2) $C_m^2$ vanishes at low frequencies for the long-wavelength modes, which approach plane waves and (3) $C_m^2$ falls abruptly down to zero above about 27 THz. This frequency corresponds in the VDOS of the present SiO$_2$ model (Fig.~\ref{dos}) to the beginning of the optic-like modes that compose the two high-frequency bands between about 27 and 40 THz~\cite{tanguy2015, damart2015,shcheblanov2015}. In-between these two limits, in the so-called main band of the VDOS, there are very rapid variations, with many modes having very low $C_m^2$ values, and a few modes having very large values. Similar spectra are obtained with other applied strains, such as shear.

Still focusing on the case of isostatic deformations, the non-affine parameter $C_m$ in Eq.~\ref{Cm} can be rewritten as:
\begin{equation}
	C_m = \sum_{i\alpha}\Xi_i^{\alpha}\frac{e_i^{\alpha}(m)}{\sqrt{m_i}},
    \label{Cm2}
\end{equation}
with $\Xi_i^\alpha$ introduced in Ref.~\onlinecite{lemaitre2006} as
\begin{equation}
\Xi_i^\alpha=\frac{F_i^\alpha}{\epsilon} =- \sum_{j\beta}D_{ij}^{\alpha \beta} R_{ij}^\beta,
   \label{Xi}
\end{equation}
where $F_i^\alpha$ is the force on atom $i$ in direction $\alpha$ induced by an affine isostatic deformation $\epsilon$ applied to the initial configuration. The atomic vector field $\Xi$ can be interpreted in two complementary ways. From the first equality in Eq. \ref{Xi}, $\Xi$ corresponds to the atomic forces resulting from an elementary affine deformation applied to the simulation cell. From the second equality, the vector $\overrightarrow{\Xi}_i$ can be interpreted as
a measure of the lack of symmetry of the atomic environment around atom $i$. This is particularly clear with a pair potential as used here since $\overrightarrow{\Xi}_i$ can be re-written in this case as:
\begin{equation}
\overrightarrow{\Xi}_i = \sum_j \phi''(R_{ij}) \overrightarrow{R}_{ij}
\label{Xi2}
\end{equation}
with $\phi(r)$ the  interatomic pair potential. If for instance, the local environment of atom $i$ is centro-symmetrical, there is for each atom $j$ at $\overrightarrow{R}_{ij}$, an atom $j'$ at $-\overrightarrow{R}_{ij}$ with an opposite contribution to $\overrightarrow{\Xi}_i$, which is therefore zero. This is true for other symmetrical environments, such as the regular tetrahedra surrounding Si atoms in SiO$_2$, since from Eq. \ref{Xi2}, $\overrightarrow{\Xi}_i$ vanishes whenever atom $i$ is at the center of gravity of its neighbors weighted by the bond strengths (measured by $\phi''$).

In the general case (Eq. \ref{Cmgeneral}), the vector field $\Xi_{\alpha\beta}$ depends on the orientation of the applied strain $\epsilon_{\alpha\beta}$:
\begin{equation}
\Xi_{\alpha\beta, i}^\kappa = -\frac{1}{2} \sum_{j}  \Big( D_{ij}^{\kappa\alpha}R_{ij}^{\beta} + D_{ij}^{\kappa\beta}R_{ij}^{\alpha} \Big).
\end{equation}
This expression cannot be simplified as above, but it retains the property of vanishing in symmetrical local environments~\cite{lemaitre2006}, justifying to quality $\Xi$ as an asymmetry vector field. Since $C_m$ is the projection of eigenvector $e(m)$ on the asymmetry field $\Xi$ (Eq. \ref{Cm2}) (we neglect here the potential effect of varying masses, which can be incorporated if needed in the definition of $\Xi$), we conclude that the modes that dissipate the most are those, which best resemble $\Xi$. This field thus encodes the information about the structural features that control harmonic dissipation when measured  with mechanical spectroscopy.

We finally note that the non-affine parameter can also be re-written in a third alternative way:
\begin{equation}
C_m = \sum_{i} \sum_{j \alpha}  \xi^{\alpha}_{j\rightarrow i}(m) R^{\alpha}_{ij},
\end{equation}
with
\begin{equation}
\xi^{\alpha}_{j\rightarrow i} = - \frac{F^\alpha_{j\rightarrow i}}{s_m} = \sum_\beta D_{ij}^{\alpha \beta} \frac{e_j^\beta(m)}{\sqrt{m_j}}
\end{equation}
where $F^\alpha_{j\rightarrow i}$ is the force on atom $i$ in direction $\alpha$ due to the displacement of atom $j$ when mode $m$ has a mass-scaled amplitude $s_m$. Therefore, $\xi^{\alpha}_{j\rightarrow i}$ measures the sensitivity on the mode amplitude of the force on atom $i$ due to atom $j$. This alternative expression shows the strong connection between dissipation and the forces induced in the glass by the eigenmodes. Such connection between force distributions and vibrational properties has recently been pointed out in the case of  hard-sphere glasses~\cite{degiuli2014b}. In silica, the most important forces are supported by the Si-O bonds and form force chains supported by the SiO$_2$ skeleton. When non-affine atomic displacements are induced by an eigenmode, the force chains adopt a specific response, reflected by $\xi^{\alpha}_{j\rightarrow i}$, which varies very rapidly from mode to mode, like $C_m$.

\subsection{Application to SiO$_2$}

\begin{figure}
 \includegraphics [width=0.5\textwidth]{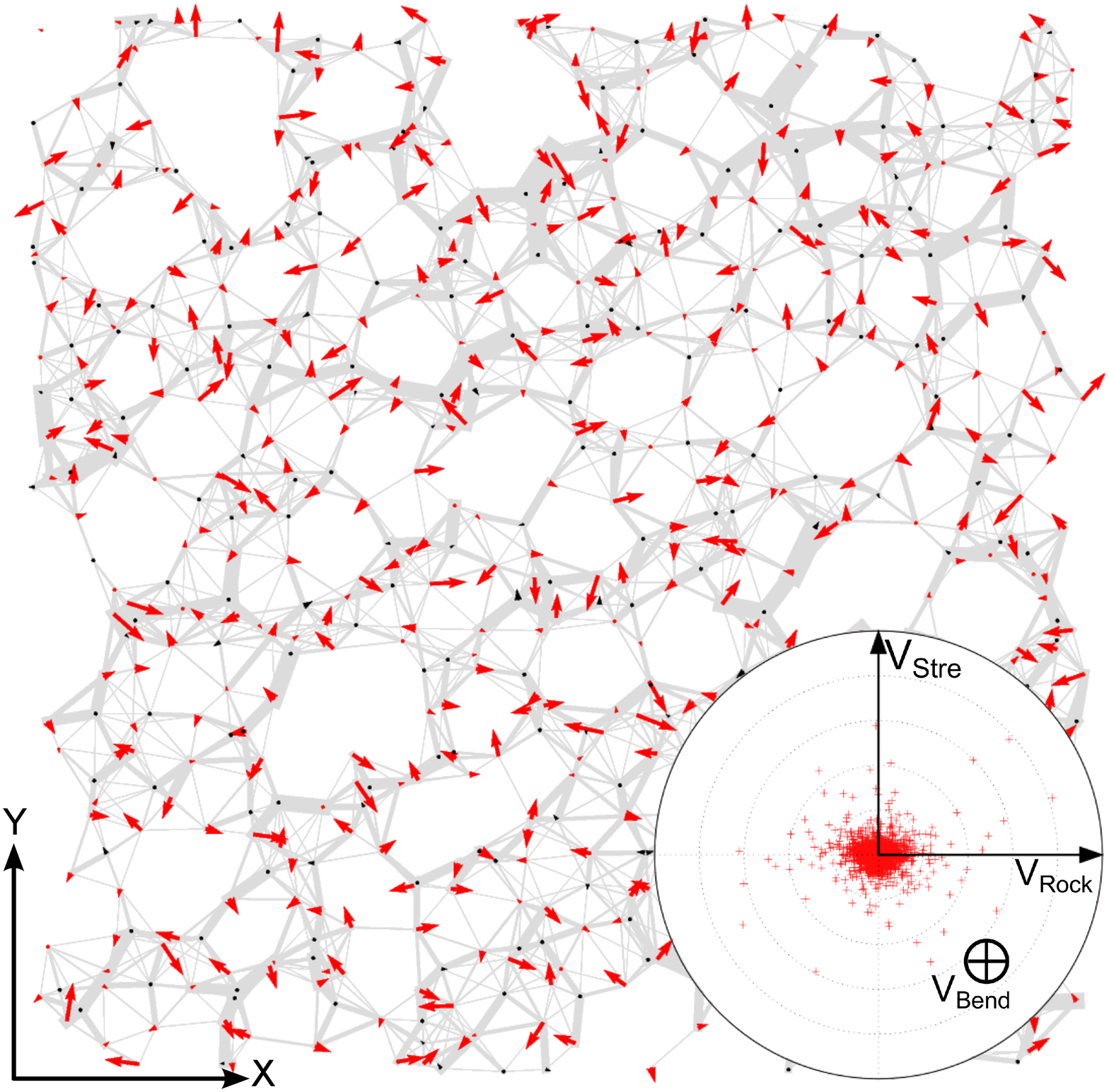}
  \caption{Asymmetry vector field $\Xi$ in the case of isostatic deformations, represented in 2D projection for an 8~$\angstrom$ slab in a SiO$_2$ sample. Black arrows are for Si atoms, red arrows for oxygen atoms. The scaled forces $\xi^{\alpha}_{j\rightarrow i}$ for the mode of maximum dissipation are shown in grey, with a width proportional to their intensity. The inset shows the stereographic projection of the $\Xi$ vectors (normalized to unity) of the O atoms in the basis formed by the rocking, stretching and bending vectors ($V_{Rock}$, $V_{Stre}$,$V_{Bend}$) of each Si-O-Si bond.}
\protect\label{symmetry}
\end{figure}

We concluded from the above discussion that inspecting the asymmetry field $\Xi$ allows to identify the structural elements responsible for harmonic dissipation. To this end, we plot in Fig.~\ref{symmetry} the $\Xi$ field for isostatic deformations in a slab of SiO$_2$.

The $\Xi$ vectors on the Si atoms (in black) are very small and hardly visible, as expected from their tetrahedral environment. Four-fold coordinated Si atoms therefore do not participate in harmonic dissipation. On the other hand, oxygen atoms are either two- or three-folded and have asymmetrical environments, resulting in finite $\Xi$ vectors (in red). Moreover, O atoms form Si-O-Si bonds and we can see in Fig.~\ref{symmetry} that in most cases, the $\Xi$ vectors point towards the inside of the Si-O-Si bond, i.e. in a direction which bends the bond. This is readily understood from Eq. \ref{Xi2}, where the O atom in a Si-O-Si bond has two Si neighbors at similar distances, resulting in a $\Xi$ vector close to the bisector vector of the Si-O-Si angle. We checked this result numerically by computing in the inset of Fig.~\ref{symmetry} the stereographic projection of the $\Xi$ vectors (normalized to unity) of the O atoms in the basis formed by the rocking, bending and stretching vectors ($V_{Rock}$, $V_{Bend}$ and $V_{Stre}$) of each Si-O-Si bond (see Fig.~\ref{dos}). Most $\Xi$ vectors are oriented along the bending vector, which by construction, is the bisector vector of the Si-O-Si angle. Therefore, in the case of isostatic deformations, bending of the Si-O-Si bonds is the main contributor to harmonic dissipation in amorphous SiO$_2$.

\begin{figure}
 \includegraphics [width=0.5\textwidth]{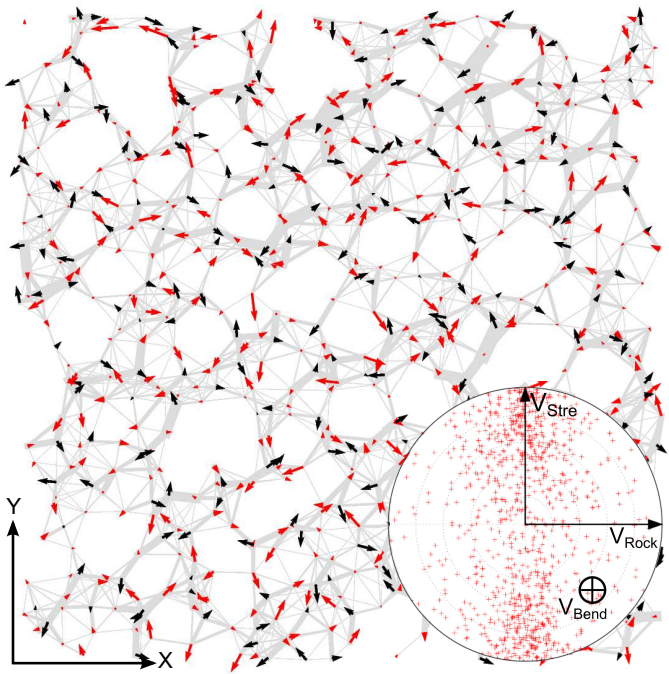}
  \caption{Asymmetry vector field $\Xi_{xy}$ and its stereographic projection for an applied shear strain $\epsilon_{xy}$ in the same 8~$\angstrom$ slab as in Fig. \ref{symmetry}, projected on the xy plane of the slab. The scale of the arrows is tripled compared to Fig. \ref{symmetry}.}
\protect\label{symmetry_shear}
\end{figure}

For a general applied strain, the $\Xi_{\alpha\beta}$ field  will remain small on the Si atoms and may take other orientations on the O atoms. As an illustration, we show in Fig. \ref{symmetry_shear} the case of simple shear. Dissipation is smaller than with isostatic deformations, as evidenced by the smaller length of the $\Xi_{xy}$ vectors (their scale was tripled compared to Fig. \ref{symmetry}). Moreover, the difference between Si and O atoms is also smaller, although O atoms still support on average larger $\Xi_{xy}$ vectors than Si atoms. Finally, the distribution of orientations of the $\Xi_{xy}$ vectors on O atoms is more spread, but the stereographic projection shows that they are predominantly oriented along $V_{stre}$. In simple shear, dissipation is therefore dominated by the stretching motion of the Si-O-Si bonds.

\section{CONCLUSION}

Mechanical spectroscopy was used to measure dissipation at high frequencies in a model SiO$_2$ glass. We have shown that the loss angle can be expressed analytically in the harmonic regime, characteristic of the high frequencies accessible to molecular dynamics simulations and inelastic x-ray scattering experiments. This analytical expression, written as a sum of damped harmonic oscillator dissipations, shows the role of the eigenmodes as energy dissipaters. The sensitivity of the stress tensor to the vibrational modes is central to understanding high frequency dissipation. Up to now however, despite its formal evidence, a more quantitative connection between eigenfrequencies and the contribution of the corresponding eigenmodes to the global stress tensor is lacking due to the complex shape of the vibrations in amorphous solids. 

We have shown that an asymmetry vector field, which depends only on the equilibrium configuration of the glass, can be used to characterize the structural features that control harmonic energy dissipation. We recover here that force asymmetries are at the origin of non-affine displacements, as discussed in the works of Lema\^itre and Maloney~\cite{lemaitre2006} and Zaccone \textit{et al.} ~\cite{zaccone2011}, and that the non-affinity of the local fields in turn are responsible for energy dissipation, as measured using mechanical spectroscopy.

In the particular case of SiO$_2$, we have shown that the deformation of the Si-O-Si bonds is the main contributor to energy dissipation. However, we should insist that since dissipation arises from the extended modes of the main band of the VDOS, dissipation is not related to the local vibration of a bond, but rather to the collective vibration of many Si-O-Si bonds. With respect to the ring structure of silica, we have seen in Fig. \ref{symmetry} that in the case of isostatic deformations, the $\Xi$ arrows point mostly towards the center of the rings, anticipating a potential connection between ring morphology and dissipation. Also, it was shown that internal friction is related to the forces induced in the SiO$_2$ skeleton by the eigenmodes, in connection with their effect on the stress tensor. This sensitivity of the stress tensor to the vibration modes of the system not only confirms that the force distribution affects vibrational properties, as pointed out in Ref. \cite{degiuli2014b}, but shows also that the components of the forces that are relevant for high-frequency internal friction are the non-trivial harmonic components induced by the displacements along the eigenmodes of the samples.

Applications of $\Xi$ as an asymmetry vector field go beyond the case of oxide glasses. For instance, we understand that in crystals, defects that break the local symmetry, such as dislocations or grain boundaries, will be sources of harmonic dissipation. Likewise, we expect that non-centrosymmetrical crystals, such as quartz, should also exhibit high-frequency dissipation, even in absence of defects.


\begin{acknowledgments}

TD and DR acknowledge support from LABEX iMUST (ANR-10-LABX-0064) of Universit\'e de Lyon (program "Investissements d'Avenir", ANR-11-IDEX-0007).

\end{acknowledgments}

\pagebreak

\appendix

\section{\label{sec:analytic}ANALYTIC EXPRESSION OF THE DISSIPATION}

We detail here the calculations leading to an analytic expression of the dissipation in Eq. \ref{analytic} in the harmonic linear response regime, assuming a Langevin thermostat. We consider here only the case of isostatic deformations. The general expressions given in the main text are obtained by a straightforward generalization of these calculations.

\subsection{Frequency-dependent pressure}

We note $r_i^{\alpha}$ the current position of atom $i$ in direction ${\alpha}$ and $R_i^{\alpha}$, its equilibrium position in the reference undeformed cell. At time $t$, the cell is compressed or stretched isostatically along the $X$, $Y$ and $Z$ directions by $\epsilon(t) = \epsilon_0\sin(\omega t)$, such that:

\begin{equation}
	r_i^{\alpha} = (1+\epsilon) R_i^{\alpha} + x_i^{\alpha},
	\label{nonaff}
\end{equation}
where $x_i^{\alpha}$ is the non-affine displacement. In the  harmonic approximation, using the expression of the energy given in Eq. \ref{energy} of the main text, the force on atom $i$ and direction $\alpha$ is given by:
\begin{equation}
	F_i^{\alpha} = - \sum_{j\beta} D_{ij}^{\alpha \beta}(r_{ij}^{\beta}-R_{ij}^{\beta})
\end{equation}
where $R_{ij}^{\alpha}$ is the equilibrium separation between atoms $i$ and $j$ in direction ${\alpha}$ in the underformed initial cell. The dynamical matrix, $\{ D_{ij}^{\alpha \beta} \}$, has the following usual symmetries $D_{ij}^{\alpha \beta} = D_{ji}^{\beta \alpha} = D_{ji}^{\alpha \beta} = D_{ij}^{\beta \alpha}$ and $\sum_i D_{ij}^{\alpha \beta} = 0$. The pressure is expressed as:
\begin{equation}
	P = \frac{1}{3V} \sum_{i{\alpha}j{\beta}}D_{ij}^{\alpha \beta}(r_{ij}^{\beta}-R_{ij}^{\beta})r_{ij}^{\alpha},
\end{equation}
where $V = L^3$ is the current volume of the cell, with $L = L_0 (1+\epsilon$). At the rather low temperatures considered here, we have checked that the kinetic pressure is negligible and will not be included in the calculations.

The pressure can be re-written as a function of the applied strain $\epsilon$ and the non-affine displacements $x_i^{\alpha}$ using Eq. \ref{nonaff}:
\begin{equation}
\begin{aligned}
P&=\frac{1}{3V}\sum_{i\alpha j\beta}D_{ij}^{\alpha \beta}\left[\epsilon R_{ij}^{\beta}+x_{ij}^{\beta}\right]\left[(1+\epsilon) R_{ij}^{\alpha}+x_{ij}^{\alpha}\right] \\
&=\frac{\epsilon(1+\epsilon)}{3V}\sum_{i\alpha j\beta}D_{ij}^{\alpha \beta}R_{ij}^{\beta} R_{ij}^{\alpha}+\frac{1}{3V}\sum_{i\alpha j\beta} D_{ij}^{\alpha \beta}x_{ij}^{\beta} x_{ij}^{\alpha} \\
&+\frac{(1+2\epsilon)}{3V}\sum_{i\alpha j\beta}D_{ij}^{\alpha \beta}R_{ij}^{\alpha}x_{ij}^{\beta}
\end{aligned}
\label{pressure11}
\end{equation}

In the linear response regime, only the first-order terms in $\epsilon$ and $x_{ij}^{\beta}$ are kept. The second term of the above expression is therefore neglected and the current volume $V$ is replaced by the reference volume $V_0$. The first term corresponds to the pressure in case of affine atomic motion, with the affine bulk modulus:
\begin{equation}
	K^\infty = - \frac{1}{3V_0}\sum_{i\alpha j\beta}D_{ij}^{\alpha \beta}R_{ij}^{\beta} R_{ij}^{\alpha}.
\end{equation}
The last term of Eq. \ref{pressure11} is the non-affine contribution, which can be re-arranged using the symmetries of $D$: $\sum_{i\alpha j\beta}D_{ij}^{\alpha \beta} R_{ij}^{\alpha}x_{ij}^{\beta} = \sum_{i\alpha j\beta}D_{ij}^{\alpha \beta} R_{ij}^{\alpha}x_{j}^{\beta}-\sum_{i\alpha j\beta}D_{ij}^{\alpha \beta} R_{ij}^{\alpha}x_{i}^{\beta}=2\sum_{i\alpha j\beta}D_{ij}^{\alpha \beta} R_{ij}^{\alpha}x_{j}^{\beta}$. The pressure is thus expressed as:

\begin{equation}
P = -K^\infty \epsilon+\frac{2}{3V_0}\sum_{i\alpha j\beta}D_{ij}^{\alpha \beta}R_{ij}^{\alpha} x_j^{\beta}.
\label{pressure0}
\end{equation}

We then project the non-affine displacements onto the normal modes of the system. To this end, we introduce mass-scaled displacements $s_j^{\beta}$ and their projections $s_m$ on the eigenmodes $e(m)$ of the system, i.e. the eigenvectors of the mass-scaled dynamical matrix $\tilde{D}$:
\begin{equation}
	s_j^{\beta} = \sqrt{m_j} x_j^{\beta} = \sum_m s_m e_j^{\beta}(m).
\end{equation}

Replacing $x_j^{\beta}$ in Eq. \ref{pressure0}, we form a mode-dependent term:

\begin{equation}
	C_m = \sum_{i\alpha j\beta}\frac{D_{ij}^{\alpha \beta}}{\sqrt{m_j}}R_{ij}^{\alpha} e_j^{\beta}(m)
    \label{CmAppen}
\end{equation}
and the final expression of the pressure is:

\begin{equation}
P = -K^\infty \epsilon + \frac{2}{3 V_0} \sum_m C_m s_m.
\end{equation}

Averaging this equation over multiple cycles and taking its Fourier transform, we obtain:

\begin{equation}
\langle P \rangle (\omega)= -K^\infty \epsilon (\omega) + \frac{2}{3 V_0} \sum_m C_m \langle s_m \rangle (\omega).
\label{P_omega}
\end{equation}

We see that $C_m$ is proportional to $\partial P/\partial s_m$ and therefore expresses the sensitivity of the pressure to the amplitude of the normal mode $m$. The properties of $C_m$ will be further explored below.

\subsection{Complex bulk modulus}

To express $\langle s_m \rangle (\omega)$, we rewrite the SSLOD equations of Eq. \ref{sllod0} in the main text as a second-order differential equation:

\begin{equation}
	m_i\ddot{r}_i^{\alpha} = m_i r_i^{\alpha} (\dot{\epsilon}^2+\ddot{\epsilon})-\sum_{j\beta}D_{ij}^{\alpha \beta}(r_{ij}^{\beta}-R_{ij}^{\beta})-m_i\gamma (\dot{r}_i^{\alpha}-r_i^{\alpha} \dot{\epsilon})+F_{th},
\end{equation}
which is written in terms of non-affine displacements, keeping only the first-order terms, as:

\begin{equation}
	m_i\ddot{x}_i^{\alpha}=-(\sum_{j\beta}D_{ij}^{\alpha \beta}R_{ij}^{\beta})\epsilon-\sum_{j\beta}D_{ij}^{\alpha \beta}x_{j}^{\beta}-m_i\gamma\dot{x}_i^{\alpha}+F_{th}.
\end{equation}

Introducing the mass-scaled coordinates $s_i^{\alpha}$:

\begin{equation}
	\ddot{s}_i^{\alpha}=-\Big(\sum_{j\beta}\frac{D_{ij}^{\alpha \beta}}{\sqrt{m_i}} R_{ij}^{\beta} \Big)\epsilon-\sum_{j\beta}\tilde{D}_{ij}^{\alpha \beta}s_{j}^{\beta}-\gamma\dot{s}_i^{\alpha}+\frac{F_{th}}{\sqrt{m_i}},
\end{equation}
which yields after projection on the eigenmodes:
\begin{equation}
	\ddot{s}_m=C_m \epsilon- \omega_m^2s_m-\gamma\dot{s}_m+F_m.
\end{equation}
Here, $F_m$ is the random force on mode $m$ and $\omega_m^2$ the eigenfrequency of mode $m$, i.e. the eigenvalue of $\tilde{D}$ corresponding to the eigenmode $e(m)$. We have also recognized that $\sum_{i\alpha j\beta}\frac{D_{ij}^{\alpha \beta}}{\sqrt{m_i}} R_{ij}^{\beta} e_i^{\alpha}(m) = -C_m$, in reference to Eq. \ref{Cm} where the minus sign comes from the exchange between indices $i$ and $j$. Averaging this equation over multiple cycles, the random force term, of zero mean, vanishes, and taking the Fourier transform, we obtain:

\begin{equation}
 	\langle s_m \rangle (\omega) = \frac{C_m}{\omega_m^2-\omega^2+i\gamma\omega} \epsilon (\omega).
\end{equation}

From Eq. \ref{P_omega}, we have:

\begin{equation}
 	\langle P \rangle (\omega) = -K^\infty \epsilon (\omega) + \frac{2}{3 V_0} \sum_m \frac{C^2_m}{\omega_m^2-\omega^2+i\gamma\omega} \langle \epsilon \rangle (\omega),
\end{equation}
resulting in the complex bulk modulus reported in the main text:

\begin{equation}
	K(\omega) = K^\infty - \frac{2}{3 V_0} \sum_m \frac{C^2_m}{\omega_m^2-\omega^2+i\gamma\omega}.
\end{equation}
\\
\\
\bibliography{bibliographie}

\begin{thebibliography}{83}
\expandafter\ifx\csname natexlab\endcsname\relax\def\natexlab#1{#1}\fi
\expandafter\ifx\csname bibnamefont\endcsname\relax
  \def\bibnamefont#1{#1}\fi
\expandafter\ifx\csname bibfnamefont\endcsname\relax
  \def\bibfnamefont#1{#1}\fi
\expandafter\ifx\csname citenamefont\endcsname\relax
  \def\citenamefont#1{#1}\fi
\expandafter\ifx\csname url\endcsname\relax
  \def\url#1{\texttt{#1}}\fi
\expandafter\ifx\csname urlprefix\endcsname\relax\def\urlprefix{URL }\fi
\providecommand{\bibinfo}[2]{#2}
\providecommand{\eprint}[2][]{\url{#2}}

\bibitem[{\citenamefont{Lifshitz and Roukes}(2000)}]{lifshitz2000}
\bibinfo{author}{\bibfnamefont{R.}~\bibnamefont{Lifshitz}} \bibnamefont{and}
  \bibinfo{author}{\bibfnamefont{M.~L.} \bibnamefont{Roukes}},
  \bibinfo{journal}{Phys. Rev. B} \textbf{\bibinfo{volume}{61}},
  \bibinfo{pages}{5600} (\bibinfo{year}{2000}).

\bibitem[{\citenamefont{Houston et~al.}(2002)\citenamefont{Houston, Photiadis,
  Marcus, Bucaro, Liu, and Vignola}}]{houston2002}
\bibinfo{author}{\bibfnamefont{B.~H.} \bibnamefont{Houston}},
  \bibinfo{author}{\bibfnamefont{D.~M.} \bibnamefont{Photiadis}},
  \bibinfo{author}{\bibfnamefont{M.~H.} \bibnamefont{Marcus}},
  \bibinfo{author}{\bibfnamefont{J.~A.} \bibnamefont{Bucaro}},
  \bibinfo{author}{\bibfnamefont{X.}~\bibnamefont{Liu}}, \bibnamefont{and}
  \bibinfo{author}{\bibfnamefont{J.~F.} \bibnamefont{Vignola}},
  \bibinfo{journal}{Appl. Phys. Lett.} \textbf{\bibinfo{volume}{80}},
  \bibinfo{pages}{1300} (\bibinfo{year}{2002}).

\bibitem[{\citenamefont{Li et~al.}(2007)\citenamefont{Li, Tang, and
  Roukes}}]{li2007}
\bibinfo{author}{\bibfnamefont{M.}~\bibnamefont{Li}},
  \bibinfo{author}{\bibfnamefont{H.~X.} \bibnamefont{Tang}}, \bibnamefont{and}
  \bibinfo{author}{\bibfnamefont{M.~L.} \bibnamefont{Roukes}},
  \bibinfo{journal}{Nat. Nanotechnol.} \textbf{\bibinfo{volume}{2}},
  \bibinfo{pages}{114} (\bibinfo{year}{2007}).

\bibitem[{\citenamefont{Saulson}(1990)}]{saulson1990}
\bibinfo{author}{\bibfnamefont{P.~R.} \bibnamefont{Saulson}},
  \bibinfo{journal}{Phys. Rev. D} \textbf{\bibinfo{volume}{42}},
  \bibinfo{pages}{2437} (\bibinfo{year}{1990}).

\bibitem[{\citenamefont{Flaminio et~al.}(2010)\citenamefont{Flaminio, Franc,
  Michel, Morgado, Pinard, and Sassolas}}]{flaminio2010}
\bibinfo{author}{\bibfnamefont{R.}~\bibnamefont{Flaminio}},
  \bibinfo{author}{\bibfnamefont{J.}~\bibnamefont{Franc}},
  \bibinfo{author}{\bibfnamefont{C.}~\bibnamefont{Michel}},
  \bibinfo{author}{\bibfnamefont{N.}~\bibnamefont{Morgado}},
  \bibinfo{author}{\bibfnamefont{L.}~\bibnamefont{Pinard}}, \bibnamefont{and}
  \bibinfo{author}{\bibfnamefont{B.}~\bibnamefont{Sassolas}},
  \bibinfo{journal}{Classical Quant. Grav.} \textbf{\bibinfo{volume}{27}},
  \bibinfo{pages}{084030} (\bibinfo{year}{2010}).

\bibitem[{\citenamefont{Granata et~al.}(2013)\citenamefont{Granata, Craig,
  Cagnoli, Carcy, Cunningham, Degallaix, Flaminio, Forest, Hart, Hennig
  et~al.}}]{granata2013}
\bibinfo{author}{\bibfnamefont{M.}~\bibnamefont{Granata}},
  \bibinfo{author}{\bibfnamefont{K.}~\bibnamefont{Craig}},
  \bibinfo{author}{\bibfnamefont{G.}~\bibnamefont{Cagnoli}},
  \bibinfo{author}{\bibfnamefont{C.}~\bibnamefont{Carcy}},
  \bibinfo{author}{\bibfnamefont{W.}~\bibnamefont{Cunningham}},
  \bibinfo{author}{\bibfnamefont{J.}~\bibnamefont{Degallaix}},
  \bibinfo{author}{\bibfnamefont{R.}~\bibnamefont{Flaminio}},
  \bibinfo{author}{\bibfnamefont{D.}~\bibnamefont{Forest}},
  \bibinfo{author}{\bibfnamefont{M.}~\bibnamefont{Hart}},
  \bibinfo{author}{\bibfnamefont{J.-S.} \bibnamefont{Hennig}},
  \bibnamefont{et~al.}, \bibinfo{journal}{Opt. Lett.}
  \textbf{\bibinfo{volume}{38}}, \bibinfo{pages}{5268} (\bibinfo{year}{2013}).

\bibitem[{\citenamefont{Seeger et~al.}(1957)\citenamefont{Seeger, Donth, and
  Pfaff}}]{seeger1957}
\bibinfo{author}{\bibfnamefont{A.}~\bibnamefont{Seeger}},
  \bibinfo{author}{\bibfnamefont{H.}~\bibnamefont{Donth}}, \bibnamefont{and}
  \bibinfo{author}{\bibfnamefont{F.}~\bibnamefont{Pfaff}},
  \bibinfo{journal}{Discuss. Faraday Soc.} \textbf{\bibinfo{volume}{23}},
  \bibinfo{pages}{19} (\bibinfo{year}{1957}).

\bibitem[{\citenamefont{Jackle}(1972)}]{jackle1972}
\bibinfo{author}{\bibfnamefont{J.}~\bibnamefont{Jackle}}, \bibinfo{journal}{Z.
  Phys.} \textbf{\bibinfo{volume}{257}}, \bibinfo{pages}{212}
  (\bibinfo{year}{1972}).

\bibitem[{\citenamefont{Phillips}(1987)}]{phillips1987}
\bibinfo{author}{\bibfnamefont{W.~A.} \bibnamefont{Phillips}},
  \bibinfo{journal}{Rep. Prog. Phys.} \textbf{\bibinfo{volume}{50}},
  \bibinfo{pages}{1657} (\bibinfo{year}{1987}).

\bibitem[{\citenamefont{Vacher et~al.}(2005)\citenamefont{Vacher, Courtens, and
  Foret}}]{vacher2005}
\bibinfo{author}{\bibfnamefont{R.}~\bibnamefont{Vacher}},
  \bibinfo{author}{\bibfnamefont{E.}~\bibnamefont{Courtens}}, \bibnamefont{and}
  \bibinfo{author}{\bibfnamefont{M.}~\bibnamefont{Foret}},
  \bibinfo{journal}{Phys. Rev. B} \textbf{\bibinfo{volume}{72}},
  \bibinfo{pages}{214205} (\bibinfo{year}{2005}).

\bibitem[{\citenamefont{Hamdan et~al.}(2014)\citenamefont{Hamdan, Trinastic,
  and Cheng}}]{hamdan2014}
\bibinfo{author}{\bibfnamefont{R.}~\bibnamefont{Hamdan}},
  \bibinfo{author}{\bibfnamefont{J.~P.} \bibnamefont{Trinastic}},
  \bibnamefont{and} \bibinfo{author}{\bibfnamefont{H.~P.} \bibnamefont{Cheng}},
  \bibinfo{journal}{J. Chem. Physics.} \textbf{\bibinfo{volume}{141}},
  \bibinfo{pages}{054501} (\bibinfo{year}{2014}).

\bibitem[{\citenamefont{Zener}(1938)}]{zener2}
\bibinfo{author}{\bibfnamefont{C.}~\bibnamefont{Zener}},
  \bibinfo{journal}{Phys. Rev.} \textbf{\bibinfo{volume}{53}},
  \bibinfo{pages}{90} (\bibinfo{year}{1938}).

\bibitem[{\citenamefont{Zener et~al.}(1938)\citenamefont{Zener, Otis, and
  Nuckolls}}]{zener3}
\bibinfo{author}{\bibfnamefont{C.}~\bibnamefont{Zener}},
  \bibinfo{author}{\bibfnamefont{W.}~\bibnamefont{Otis}}, \bibnamefont{and}
  \bibinfo{author}{\bibfnamefont{R.}~\bibnamefont{Nuckolls}},
  \bibinfo{journal}{Phys. Rev.} \textbf{\bibinfo{volume}{53}},
  \bibinfo{pages}{100} (\bibinfo{year}{1938}).

\bibitem[{\citenamefont{Fabian and Allen}(1999)}]{fabian1999}
\bibinfo{author}{\bibfnamefont{J.}~\bibnamefont{Fabian}} \bibnamefont{and}
  \bibinfo{author}{\bibfnamefont{P.~B.} \bibnamefont{Allen}},
  \bibinfo{journal}{Phys. Rev. Lett.} \textbf{\bibinfo{volume}{82}},
  \bibinfo{pages}{1478} (\bibinfo{year}{1999}).

\bibitem[{\citenamefont{Kunal and Aluru}(2011)}]{kunal2011}
\bibinfo{author}{\bibfnamefont{K.}~\bibnamefont{Kunal}} \bibnamefont{and}
  \bibinfo{author}{\bibfnamefont{N.~R.} \bibnamefont{Aluru}},
  \bibinfo{journal}{Phys. Rev. B} \textbf{\bibinfo{volume}{84}},
  \bibinfo{pages}{245450} (\bibinfo{year}{2011}).

\bibitem[{\citenamefont{Carini et~al.}(2014)\citenamefont{Carini, Carini,
  D'Angelo, Fioretto, and Tripodo}}]{carini2014}
\bibinfo{author}{\bibfnamefont{G.}~\bibnamefont{Carini}},
  \bibinfo{author}{\bibfnamefont{G.}~\bibnamefont{Carini}},
  \bibinfo{author}{\bibfnamefont{G.}~\bibnamefont{D'Angelo}},
  \bibinfo{author}{\bibfnamefont{D.}~\bibnamefont{Fioretto}}, \bibnamefont{and}
  \bibinfo{author}{\bibfnamefont{G.}~\bibnamefont{Tripodo}},
  \bibinfo{journal}{Phys. Rev. B} \textbf{\bibinfo{volume}{90}},
  \bibinfo{pages}{140204(R)} (\bibinfo{year}{2014}).

\bibitem[{\citenamefont{Ruocco et~al.}(1999)\citenamefont{Ruocco, Sette,
  Di~Leonardo, Fioretto, Krisch, Lorenzen, Masciovecchio, Monaco, Pignon, and
  Scopigno}}]{ruocco1999}
\bibinfo{author}{\bibfnamefont{G.}~\bibnamefont{Ruocco}},
  \bibinfo{author}{\bibfnamefont{F.}~\bibnamefont{Sette}},
  \bibinfo{author}{\bibfnamefont{R.}~\bibnamefont{Di~Leonardo}},
  \bibinfo{author}{\bibfnamefont{D.}~\bibnamefont{Fioretto}},
  \bibinfo{author}{\bibfnamefont{M.}~\bibnamefont{Krisch}},
  \bibinfo{author}{\bibfnamefont{M.}~\bibnamefont{Lorenzen}},
  \bibinfo{author}{\bibfnamefont{C.}~\bibnamefont{Masciovecchio}},
  \bibinfo{author}{\bibfnamefont{G.}~\bibnamefont{Monaco}},
  \bibinfo{author}{\bibfnamefont{F.}~\bibnamefont{Pignon}}, \bibnamefont{and}
  \bibinfo{author}{\bibfnamefont{T.}~\bibnamefont{Scopigno}},
  \bibinfo{journal}{Phys. Rev. Lett.} \textbf{\bibinfo{volume}{83}},
  \bibinfo{pages}{5583} (\bibinfo{year}{1999}).

\bibitem[{\citenamefont{Ganter and Schirmacher}(2010)}]{ganter2010}
\bibinfo{author}{\bibfnamefont{C.}~\bibnamefont{Ganter}} \bibnamefont{and}
  \bibinfo{author}{\bibfnamefont{W.}~\bibnamefont{Schirmacher}},
  \bibinfo{journal}{Phys. Rev. B} \textbf{\bibinfo{volume}{82}},
  \bibinfo{pages}{094205} (\bibinfo{year}{2010}).

\bibitem[{\citenamefont{Liang and Keblinski}(2016)}]{liang2016}
\bibinfo{author}{\bibfnamefont{Z.}~\bibnamefont{Liang}} \bibnamefont{and}
  \bibinfo{author}{\bibfnamefont{P.}~\bibnamefont{Keblinski}},
  \bibinfo{journal}{Phys. Rev. B} \textbf{\bibinfo{volume}{93}},
  \bibinfo{pages}{054205} (\bibinfo{year}{2016}).

\bibitem[{\citenamefont{Baldi et~al.}(2016)\citenamefont{Baldi, Giordano, Ruta,
  and Monaco}}]{baldi2016}
\bibinfo{author}{\bibfnamefont{G.}~\bibnamefont{Baldi}},
  \bibinfo{author}{\bibfnamefont{V.~M.} \bibnamefont{Giordano}},
  \bibinfo{author}{\bibfnamefont{B.}~\bibnamefont{Ruta}}, \bibnamefont{and}
  \bibinfo{author}{\bibfnamefont{G.}~\bibnamefont{Monaco}},
  \bibinfo{journal}{Phys. Rev. B} \textbf{\bibinfo{volume}{93}},
  \bibinfo{pages}{144204} (\bibinfo{year}{2016}).

\bibitem[{\citenamefont{Masciovecchio et~al.}(2006)\citenamefont{Masciovecchio,
  Baldi, Caponi, Comez, Di~Fonzo, Fioretto, Fontana, Gessini, Santucci, and
  Sette}}]{masciovecchio2006}
\bibinfo{author}{\bibfnamefont{C.}~\bibnamefont{Masciovecchio}},
  \bibinfo{author}{\bibfnamefont{G.}~\bibnamefont{Baldi}},
  \bibinfo{author}{\bibfnamefont{S.}~\bibnamefont{Caponi}},
  \bibinfo{author}{\bibfnamefont{L.}~\bibnamefont{Comez}},
  \bibinfo{author}{\bibfnamefont{S.}~\bibnamefont{Di~Fonzo}},
  \bibinfo{author}{\bibfnamefont{D.}~\bibnamefont{Fioretto}},
  \bibinfo{author}{\bibfnamefont{A.}~\bibnamefont{Fontana}},
  \bibinfo{author}{\bibfnamefont{A.}~\bibnamefont{Gessini}},
  \bibinfo{author}{\bibfnamefont{S.~C.} \bibnamefont{Santucci}},
  \bibnamefont{and} \bibinfo{author}{\bibfnamefont{F.}~\bibnamefont{Sette}},
  \bibinfo{journal}{Phys. Rev. Lett.} \textbf{\bibinfo{volume}{97}},
  \bibinfo{pages}{035501} (\bibinfo{year}{2006}).

\bibitem[{\citenamefont{Baldi et~al.}(2014)\citenamefont{Baldi, Giordano, Ruta,
  Dal~Maschio, Fontana, and Monaco}}]{baldi2014}
\bibinfo{author}{\bibfnamefont{G.}~\bibnamefont{Baldi}},
  \bibinfo{author}{\bibfnamefont{V.~M.} \bibnamefont{Giordano}},
  \bibinfo{author}{\bibfnamefont{B.}~\bibnamefont{Ruta}},
  \bibinfo{author}{\bibfnamefont{R.}~\bibnamefont{Dal~Maschio}},
  \bibinfo{author}{\bibfnamefont{A.}~\bibnamefont{Fontana}}, \bibnamefont{and}
  \bibinfo{author}{\bibfnamefont{G.}~\bibnamefont{Monaco}},
  \bibinfo{journal}{Phys. Rev. Lett.} \textbf{\bibinfo{volume}{112}},
  \bibinfo{pages}{125502} (\bibinfo{year}{2014}).

\bibitem[{\citenamefont{Hansen and McDonald}(2006)}]{hansen2006}
\bibinfo{author}{\bibfnamefont{J.-P.} \bibnamefont{Hansen}} \bibnamefont{and}
  \bibinfo{author}{\bibfnamefont{I.~R.} \bibnamefont{McDonald}},
  \emph{\bibinfo{title}{Theory of simple liquids}}
  (\bibinfo{publisher}{Elsevier Academic Press}, \bibinfo{year}{2006}).

\bibitem[{\citenamefont{Foret et~al.}(1996)\citenamefont{Foret, Courtens,
  Vacher, and Suck}}]{foret1996}
\bibinfo{author}{\bibfnamefont{M.}~\bibnamefont{Foret}},
  \bibinfo{author}{\bibfnamefont{E.}~\bibnamefont{Courtens}},
  \bibinfo{author}{\bibfnamefont{R.}~\bibnamefont{Vacher}}, \bibnamefont{and}
  \bibinfo{author}{\bibfnamefont{J.-B.} \bibnamefont{Suck}},
  \bibinfo{journal}{Phys. Rev. Lett.} \textbf{\bibinfo{volume}{77}},
  \bibinfo{pages}{3831} (\bibinfo{year}{1996}).

\bibitem[{\citenamefont{Ruocco and Sette}(2001)}]{ruocco2001}
\bibinfo{author}{\bibfnamefont{G.}~\bibnamefont{Ruocco}} \bibnamefont{and}
  \bibinfo{author}{\bibfnamefont{F.}~\bibnamefont{Sette}}, \bibinfo{journal}{J.
  Phys-Condens. Mat.} \textbf{\bibinfo{volume}{13}}, \bibinfo{pages}{9141}
  (\bibinfo{year}{2001}).

\bibitem[{\citenamefont{Benassi et~al.}(1996)\citenamefont{Benassi, Krisch,
  Masciovecchio, Mazzacurati, Monaco, Ruocco, Sette, and
  Verbeni}}]{benassi1996}
\bibinfo{author}{\bibfnamefont{P.}~\bibnamefont{Benassi}},
  \bibinfo{author}{\bibfnamefont{M.}~\bibnamefont{Krisch}},
  \bibinfo{author}{\bibfnamefont{C.}~\bibnamefont{Masciovecchio}},
  \bibinfo{author}{\bibfnamefont{V.}~\bibnamefont{Mazzacurati}},
  \bibinfo{author}{\bibfnamefont{G.}~\bibnamefont{Monaco}},
  \bibinfo{author}{\bibfnamefont{G.}~\bibnamefont{Ruocco}},
  \bibinfo{author}{\bibfnamefont{F.}~\bibnamefont{Sette}}, \bibnamefont{and}
  \bibinfo{author}{\bibfnamefont{R.}~\bibnamefont{Verbeni}},
  \bibinfo{journal}{Phys. Rev. Lett.} \textbf{\bibinfo{volume}{77}},
  \bibinfo{pages}{3835} (\bibinfo{year}{1996}).

\bibitem[{\citenamefont{Devos et~al.}(2008)\citenamefont{Devos, Foret,
  Ayrinhac, Emery, and Ruffl\'{e}}}]{devos2008}
\bibinfo{author}{\bibfnamefont{A.}~\bibnamefont{Devos}},
  \bibinfo{author}{\bibfnamefont{M.}~\bibnamefont{Foret}},
  \bibinfo{author}{\bibfnamefont{S.}~\bibnamefont{Ayrinhac}},
  \bibinfo{author}{\bibfnamefont{P.}~\bibnamefont{Emery}}, \bibnamefont{and}
  \bibinfo{author}{\bibfnamefont{B.}~\bibnamefont{Ruffl\'{e}}},
  \bibinfo{journal}{Phys. Rev. B} \textbf{\bibinfo{volume}{77}},
  \bibinfo{pages}{100201(R)} (\bibinfo{year}{2008}).

\bibitem[{\citenamefont{Grest et~al.}(1982)\citenamefont{Grest, Nagel, and
  Rahman}}]{grest1982}
\bibinfo{author}{\bibfnamefont{G.~S.} \bibnamefont{Grest}},
  \bibinfo{author}{\bibfnamefont{S.~R.} \bibnamefont{Nagel}}, \bibnamefont{and}
  \bibinfo{author}{\bibfnamefont{A.}~\bibnamefont{Rahman}},
  \bibinfo{journal}{Phys. Rev. Lett.} \textbf{\bibinfo{volume}{49}},
  \bibinfo{pages}{1271} (\bibinfo{year}{1982}).

\bibitem[{\citenamefont{Taraskin and
  Elliott}(1997{\natexlab{a}})}]{taraskin1997}
\bibinfo{author}{\bibfnamefont{S.~N.} \bibnamefont{Taraskin}} \bibnamefont{and}
  \bibinfo{author}{\bibfnamefont{S.~R.} \bibnamefont{Elliott}},
  \bibinfo{journal}{Phys. Rev. B} \textbf{\bibinfo{volume}{56}},
  \bibinfo{pages}{8605} (\bibinfo{year}{1997}{\natexlab{a}}).

\bibitem[{\citenamefont{Dell'Anna et~al.}(1998)\citenamefont{Dell'Anna, Ruocco,
  Sampoli, and Viliani}}]{dellanna1998}
\bibinfo{author}{\bibfnamefont{R.}~\bibnamefont{Dell'Anna}},
  \bibinfo{author}{\bibfnamefont{G.}~\bibnamefont{Ruocco}},
  \bibinfo{author}{\bibfnamefont{M.}~\bibnamefont{Sampoli}}, \bibnamefont{and}
  \bibinfo{author}{\bibfnamefont{G.}~\bibnamefont{Viliani}},
  \bibinfo{journal}{Phys. Rev. Lett.} \textbf{\bibinfo{volume}{80}},
  \bibinfo{pages}{1236} (\bibinfo{year}{1998}).

\bibitem[{\citenamefont{Ruocco et~al.}(2000)\citenamefont{Ruocco, Sette,
  Di~Leonardo, Monaco, Sampoli, Scopigno, and Viliani}}]{ruocco2000}
\bibinfo{author}{\bibfnamefont{G.}~\bibnamefont{Ruocco}},
  \bibinfo{author}{\bibfnamefont{F.}~\bibnamefont{Sette}},
  \bibinfo{author}{\bibfnamefont{R.}~\bibnamefont{Di~Leonardo}},
  \bibinfo{author}{\bibfnamefont{G.}~\bibnamefont{Monaco}},
  \bibinfo{author}{\bibfnamefont{M.}~\bibnamefont{Sampoli}},
  \bibinfo{author}{\bibfnamefont{T.}~\bibnamefont{Scopigno}}, \bibnamefont{and}
  \bibinfo{author}{\bibfnamefont{G.}~\bibnamefont{Viliani}},
  \bibinfo{journal}{Phys. Rev. Lett.} \textbf{\bibinfo{volume}{84}},
  \bibinfo{pages}{5788} (\bibinfo{year}{2000}).

\bibitem[{\citenamefont{Shintani and Tanaka}(2008)}]{shintani2008}
\bibinfo{author}{\bibfnamefont{H.}~\bibnamefont{Shintani}} \bibnamefont{and}
  \bibinfo{author}{\bibfnamefont{H.}~\bibnamefont{Tanaka}},
  \bibinfo{journal}{Nat. Mater.} \textbf{\bibinfo{volume}{7}},
  \bibinfo{pages}{870} (\bibinfo{year}{2008}).

\bibitem[{\citenamefont{Monaco and Mossa}(2009)}]{monaco2009}
\bibinfo{author}{\bibfnamefont{G.}~\bibnamefont{Monaco}} \bibnamefont{and}
  \bibinfo{author}{\bibfnamefont{S.}~\bibnamefont{Mossa}}, \bibinfo{journal}{P.
  Natl. Acad. Sci.} \textbf{\bibinfo{volume}{106}}, \bibinfo{pages}{16907}
  (\bibinfo{year}{2009}).

\bibitem[{\citenamefont{Marruzzo et~al.}(2013)\citenamefont{Marruzzo,
  Schirmacher, Fratalocchi, and Ruocco}}]{marruzzo2013}
\bibinfo{author}{\bibfnamefont{A.}~\bibnamefont{Marruzzo}},
  \bibinfo{author}{\bibfnamefont{W.}~\bibnamefont{Schirmacher}},
  \bibinfo{author}{\bibfnamefont{A.}~\bibnamefont{Fratalocchi}},
  \bibnamefont{and} \bibinfo{author}{\bibfnamefont{G.}~\bibnamefont{Ruocco}},
  \bibinfo{journal}{Sci. Rep.} \textbf{\bibinfo{volume}{3}},
  \bibinfo{pages}{1407} (\bibinfo{year}{2013}).

\bibitem[{\citenamefont{Beltukov et~al.}(2013)\citenamefont{Beltukov, Kozub,
  and Parshin}}]{beltukov2013}
\bibinfo{author}{\bibfnamefont{Y.~M.} \bibnamefont{Beltukov}},
  \bibinfo{author}{\bibfnamefont{V.~I.} \bibnamefont{Kozub}}, \bibnamefont{and}
  \bibinfo{author}{\bibfnamefont{D.~A.} \bibnamefont{Parshin}},
  \bibinfo{journal}{Phys. Rev. B} \textbf{\bibinfo{volume}{87}},
  \bibinfo{pages}{134203} (\bibinfo{year}{2013}).

\bibitem[{\citenamefont{Beltukov et~al.}(2016)\citenamefont{Beltukov, Fusco,
  Parshin, and Tanguy}}]{beltukov2016}
\bibinfo{author}{\bibfnamefont{Y.~M.} \bibnamefont{Beltukov}},
  \bibinfo{author}{\bibfnamefont{C.}~\bibnamefont{Fusco}},
  \bibinfo{author}{\bibfnamefont{D.~A.} \bibnamefont{Parshin}},
  \bibnamefont{and} \bibinfo{author}{\bibfnamefont{A.}~\bibnamefont{Tanguy}},
  \bibinfo{journal}{Phys. Rev. E} \textbf{\bibinfo{volume}{93}},
  \bibinfo{pages}{023006} (\bibinfo{year}{2016}).

\bibitem[{\citenamefont{Horbach et~al.}(2001)\citenamefont{Horbach, Kob, and
  Binder}}]{horbach2001}
\bibinfo{author}{\bibfnamefont{J.}~\bibnamefont{Horbach}},
  \bibinfo{author}{\bibfnamefont{W.}~\bibnamefont{Kob}}, \bibnamefont{and}
  \bibinfo{author}{\bibfnamefont{K.}~\bibnamefont{Binder}},
  \bibinfo{journal}{Eur. Phys. J. B} \textbf{\bibinfo{volume}{19}},
  \bibinfo{pages}{531} (\bibinfo{year}{2001}).

\bibitem[{\citenamefont{Baldi et~al.}(2010)\citenamefont{Baldi, Giordano,
  Monaco, and Ruta}}]{baldi2010}
\bibinfo{author}{\bibfnamefont{G.}~\bibnamefont{Baldi}},
  \bibinfo{author}{\bibfnamefont{V.~M.} \bibnamefont{Giordano}},
  \bibinfo{author}{\bibfnamefont{G.}~\bibnamefont{Monaco}}, \bibnamefont{and}
  \bibinfo{author}{\bibfnamefont{B.}~\bibnamefont{Ruta}},
  \bibinfo{journal}{Phys. Rev. Lett.} \textbf{\bibinfo{volume}{104}},
  \bibinfo{pages}{195501} (\bibinfo{year}{2010}).

\bibitem[{\citenamefont{Parke}(1966)}]{parke1966}
\bibinfo{author}{\bibfnamefont{S.}~\bibnamefont{Parke}}, \bibinfo{journal}{Br.
  J. Appl. Phys.} \textbf{\bibinfo{volume}{17}}, \bibinfo{pages}{271}
  (\bibinfo{year}{1966}).

\bibitem[{\citenamefont{Elliott}(1984)}]{elliott1984}
\bibinfo{author}{\bibfnamefont{S.~R.} \bibnamefont{Elliott}},
  \emph{\bibinfo{title}{Physics of amorphous materials}}
  (\bibinfo{publisher}{Longman}, \bibinfo{year}{1984}).

\bibitem[{\citenamefont{Buchenau et~al.}(1986)\citenamefont{Buchenau, Prager,
  N\"{u}cker, Dianoux, Ahmad, and Phillips}}]{buchenau1986}
\bibinfo{author}{\bibfnamefont{U.}~\bibnamefont{Buchenau}},
  \bibinfo{author}{\bibfnamefont{M.}~\bibnamefont{Prager}},
  \bibinfo{author}{\bibfnamefont{N.}~\bibnamefont{N\"{u}cker}},
  \bibinfo{author}{\bibfnamefont{A.~J.} \bibnamefont{Dianoux}},
  \bibinfo{author}{\bibfnamefont{N.}~\bibnamefont{Ahmad}}, \bibnamefont{and}
  \bibinfo{author}{\bibfnamefont{W.~A.} \bibnamefont{Phillips}},
  \bibinfo{journal}{Phys. Rev. B} \textbf{\bibinfo{volume}{34}},
  \bibinfo{pages}{5665} (\bibinfo{year}{1986}).

\bibitem[{\citenamefont{Binder and Kob}(2005)}]{binder2005}
\bibinfo{author}{\bibfnamefont{K.}~\bibnamefont{Binder}} \bibnamefont{and}
  \bibinfo{author}{\bibfnamefont{W.}~\bibnamefont{Kob}},
  \emph{\bibinfo{title}{Glassy materials and disordered solids}}
  (\bibinfo{publisher}{World Scientific Pub. Co.}, \bibinfo{year}{2005}).

\bibitem[{\citenamefont{Duval et~al.}(2007)\citenamefont{Duval, Mermet, and
  Saviot}}]{duval2007}
\bibinfo{author}{\bibfnamefont{E.}~\bibnamefont{Duval}},
  \bibinfo{author}{\bibfnamefont{A.}~\bibnamefont{Mermet}}, \bibnamefont{and}
  \bibinfo{author}{\bibfnamefont{L.}~\bibnamefont{Saviot}},
  \bibinfo{journal}{Phys. Rev. B} \textbf{\bibinfo{volume}{75}},
  \bibinfo{pages}{024201} (\bibinfo{year}{2007}).

\bibitem[{\citenamefont{Ruffl\'{e} et~al.}(2006)\citenamefont{Ruffl\'{e},
  Guimbreti\`{e}re, Courtens, Vacher, and Monaco}}]{ruffle2006}
\bibinfo{author}{\bibfnamefont{B.}~\bibnamefont{Ruffl\'{e}}},
  \bibinfo{author}{\bibfnamefont{G.}~\bibnamefont{Guimbreti\`{e}re}},
  \bibinfo{author}{\bibfnamefont{E.}~\bibnamefont{Courtens}},
  \bibinfo{author}{\bibfnamefont{R.}~\bibnamefont{Vacher}}, \bibnamefont{and}
  \bibinfo{author}{\bibfnamefont{G.}~\bibnamefont{Monaco}},
  \bibinfo{journal}{Phys. Rev. Lett.} \textbf{\bibinfo{volume}{96}},
  \bibinfo{pages}{045502} (\bibinfo{year}{2006}).

\bibitem[{\citenamefont{Ferrante et~al.}(2013)\citenamefont{Ferrante,
  Pontecorvo, Cerullo, Chiasera, Ruocco, Schirmacher, and
  Scopigno}}]{ferrante2013}
\bibinfo{author}{\bibfnamefont{C.}~\bibnamefont{Ferrante}},
  \bibinfo{author}{\bibfnamefont{E.}~\bibnamefont{Pontecorvo}},
  \bibinfo{author}{\bibfnamefont{G.}~\bibnamefont{Cerullo}},
  \bibinfo{author}{\bibfnamefont{A.}~\bibnamefont{Chiasera}},
  \bibinfo{author}{\bibfnamefont{G.}~\bibnamefont{Ruocco}},
  \bibinfo{author}{\bibfnamefont{W.}~\bibnamefont{Schirmacher}},
  \bibnamefont{and} \bibinfo{author}{\bibfnamefont{T.}~\bibnamefont{Scopigno}},
  \bibinfo{journal}{Nat. Commun.} \textbf{\bibinfo{volume}{4}},
  \bibinfo{pages}{1793} (\bibinfo{year}{2013}).

\bibitem[{\citenamefont{Schirmacher et~al.}(1998)\citenamefont{Schirmacher,
  Diezemann, and Ganter}}]{schirmacher1998}
\bibinfo{author}{\bibfnamefont{W.}~\bibnamefont{Schirmacher}},
  \bibinfo{author}{\bibfnamefont{G.}~\bibnamefont{Diezemann}},
  \bibnamefont{and} \bibinfo{author}{\bibfnamefont{C.}~\bibnamefont{Ganter}},
  \bibinfo{journal}{Phys. Rev. Lett.} \textbf{\bibinfo{volume}{81}},
  \bibinfo{pages}{136} (\bibinfo{year}{1998}).

\bibitem[{\citenamefont{Grigera et~al.}(2001)\citenamefont{Grigera,
  Martin-Mayor, Parisi, and Verrocchio}}]{grigera2001}
\bibinfo{author}{\bibfnamefont{T.~S.} \bibnamefont{Grigera}},
  \bibinfo{author}{\bibfnamefont{V.}~\bibnamefont{Martin-Mayor}},
  \bibinfo{author}{\bibfnamefont{G.}~\bibnamefont{Parisi}}, \bibnamefont{and}
  \bibinfo{author}{\bibfnamefont{P.}~\bibnamefont{Verrocchio}},
  \bibinfo{journal}{Phys. Rev. Lett.} \textbf{\bibinfo{volume}{87}},
  \bibinfo{pages}{085502} (\bibinfo{year}{2001}).

\bibitem[{\citenamefont{DeGiuli
  et~al.}(2014{\natexlab{a}})\citenamefont{DeGiuli, Laversanne-Finot,
  D\"{u}ring, Lerner, and Wyart}}]{degiuli2014}
\bibinfo{author}{\bibfnamefont{E.}~\bibnamefont{DeGiuli}},
  \bibinfo{author}{\bibfnamefont{A.}~\bibnamefont{Laversanne-Finot}},
  \bibinfo{author}{\bibfnamefont{G.}~\bibnamefont{D\"{u}ring}},
  \bibinfo{author}{\bibfnamefont{E.}~\bibnamefont{Lerner}}, \bibnamefont{and}
  \bibinfo{author}{\bibfnamefont{M.}~\bibnamefont{Wyart}},
  \bibinfo{journal}{Soft Matter} \textbf{\bibinfo{volume}{10}},
  \bibinfo{pages}{5628} (\bibinfo{year}{2014}{\natexlab{a}}).

\bibitem[{\citenamefont{Gelin et~al.}(2016)\citenamefont{Gelin, Tanaka, and
  Lema\^{i}tre}}]{gelin2016}
\bibinfo{author}{\bibfnamefont{S.}~\bibnamefont{Gelin}},
  \bibinfo{author}{\bibfnamefont{H.}~\bibnamefont{Tanaka}}, \bibnamefont{and}
  \bibinfo{author}{\bibfnamefont{A.}~\bibnamefont{Lema\^{i}tre}},
  \bibinfo{journal}{Nat. Mater.} \textbf{\bibinfo{volume}{4736}},
  \bibinfo{pages}{1177–1181} (\bibinfo{year}{2016}).

\bibitem[{\citenamefont{Tanguy et~al.}(2002)\citenamefont{Tanguy, Wittmer,
  Leonforte, and Barrat}}]{tanguy2002}
\bibinfo{author}{\bibfnamefont{A.}~\bibnamefont{Tanguy}},
  \bibinfo{author}{\bibfnamefont{J.~P.} \bibnamefont{Wittmer}},
  \bibinfo{author}{\bibfnamefont{F.}~\bibnamefont{Leonforte}},
  \bibnamefont{and} \bibinfo{author}{\bibfnamefont{J.-L.}
  \bibnamefont{Barrat}}, \bibinfo{journal}{Phys. Rev. B}
  \textbf{\bibinfo{volume}{66}}, \bibinfo{pages}{174205}
  (\bibinfo{year}{2002}).

\bibitem[{\citenamefont{Tsamados et~al.}(2009)\citenamefont{Tsamados, Tanguy,
  Goldenberg, and Barrat}}]{tsamados2009}
\bibinfo{author}{\bibfnamefont{M.}~\bibnamefont{Tsamados}},
  \bibinfo{author}{\bibfnamefont{A.}~\bibnamefont{Tanguy}},
  \bibinfo{author}{\bibfnamefont{C.}~\bibnamefont{Goldenberg}},
  \bibnamefont{and} \bibinfo{author}{\bibfnamefont{J.-L.}
  \bibnamefont{Barrat}}, \bibinfo{journal}{Phys. Rev. E}
  \textbf{\bibinfo{volume}{80}}, \bibinfo{pages}{026112}
  (\bibinfo{year}{2009}).

\bibitem[{\citenamefont{Wagner et~al.}(2011)\citenamefont{Wagner, Bedorf,
  Küchemann, Schwabe, Zhang, Arnold, and Samwer}}]{wagner2011}
\bibinfo{author}{\bibfnamefont{H.}~\bibnamefont{Wagner}},
  \bibinfo{author}{\bibfnamefont{D.}~\bibnamefont{Bedorf}},
  \bibinfo{author}{\bibfnamefont{S.}~\bibnamefont{Küchemann}},
  \bibinfo{author}{\bibfnamefont{M.}~\bibnamefont{Schwabe}},
  \bibinfo{author}{\bibfnamefont{B.}~\bibnamefont{Zhang}},
  \bibinfo{author}{\bibfnamefont{W.}~\bibnamefont{Arnold}}, \bibnamefont{and}
  \bibinfo{author}{\bibfnamefont{K.}~\bibnamefont{Samwer}},
  \bibinfo{journal}{Nat. Mater.} \textbf{\bibinfo{volume}{10}},
  \bibinfo{pages}{439} (\bibinfo{year}{2011}).

\bibitem[{\citenamefont{Tanguy}(2015)}]{tanguy2015}
\bibinfo{author}{\bibfnamefont{A.}~\bibnamefont{Tanguy}},
  \bibinfo{journal}{JOM} \textbf{\bibinfo{volume}{67}}, \bibinfo{pages}{1832}
  (\bibinfo{year}{2015}).

\bibitem[{\citenamefont{Ioffe and Regel}(1960)}]{ioffe1960}
\bibinfo{author}{\bibfnamefont{A.~F.} \bibnamefont{Ioffe}} \bibnamefont{and}
  \bibinfo{author}{\bibfnamefont{A.~R.} \bibnamefont{Regel}},
  \bibinfo{journal}{Prog. Semicond.} \textbf{\bibinfo{volume}{4}},
  \bibinfo{pages}{89} (\bibinfo{year}{1960}).

\bibitem[{\citenamefont{Allen et~al.}(1999)\citenamefont{Allen, Feldman,
  Fabian, and Wooten}}]{allen1999}
\bibinfo{author}{\bibfnamefont{P.~B.} \bibnamefont{Allen}},
  \bibinfo{author}{\bibfnamefont{J.~L.} \bibnamefont{Feldman}},
  \bibinfo{author}{\bibfnamefont{J.}~\bibnamefont{Fabian}}, \bibnamefont{and}
  \bibinfo{author}{\bibfnamefont{F.}~\bibnamefont{Wooten}},
  \bibinfo{journal}{Philos. Mag. B} \textbf{\bibinfo{volume}{79}},
  \bibinfo{pages}{1715} (\bibinfo{year}{1999}).

\bibitem[{\citenamefont{Buchenau}(2014)}]{buchenau2014}
\bibinfo{author}{\bibfnamefont{U.}~\bibnamefont{Buchenau}},
  \bibinfo{journal}{Phys. Rev. E} \textbf{\bibinfo{volume}{90}},
  \bibinfo{pages}{062319} (\bibinfo{year}{2014}).

\bibitem[{\citenamefont{Qiao et~al.}(2013)\citenamefont{Qiao, Pelletier, and
  Casalini}}]{qiao2013}
\bibinfo{author}{\bibfnamefont{J.}~\bibnamefont{Qiao}},
  \bibinfo{author}{\bibfnamefont{J.-M.} \bibnamefont{Pelletier}},
  \bibnamefont{and} \bibinfo{author}{\bibfnamefont{R.}~\bibnamefont{Casalini}},
  \bibinfo{journal}{J. Phys. Chem. B} \textbf{\bibinfo{volume}{117}},
  \bibinfo{pages}{13658} (\bibinfo{year}{2013}).

\bibitem[{\citenamefont{Liu et~al.}(2015)\citenamefont{Liu, Pineda, and
  Crespo}}]{liu2015}
\bibinfo{author}{\bibfnamefont{C.}~\bibnamefont{Liu}},
  \bibinfo{author}{\bibfnamefont{E.}~\bibnamefont{Pineda}}, \bibnamefont{and}
  \bibinfo{author}{\bibfnamefont{D.}~\bibnamefont{Crespo}},
  \bibinfo{journal}{Metals} \textbf{\bibinfo{volume}{5}}, \bibinfo{pages}{1073}
  (\bibinfo{year}{2015}).

\bibitem[{\citenamefont{Rubinstein and Colby}(2003)}]{rubinstein2003}
\bibinfo{author}{\bibfnamefont{M.}~\bibnamefont{Rubinstein}} \bibnamefont{and}
  \bibinfo{author}{\bibfnamefont{R.~H.} \bibnamefont{Colby}},
  \emph{\bibinfo{title}{Polymer physics}} (\bibinfo{publisher}{Oxford
  University Press}, \bibinfo{year}{2003}).

\bibitem[{\citenamefont{Coussot}(2005)}]{coussot2005}
\bibinfo{author}{\bibfnamefont{P.}~\bibnamefont{Coussot}},
  \emph{\bibinfo{title}{Rheometry of pastes, suspensions, and granular
  materials}} (\bibinfo{publisher}{Wiley}, \bibinfo{year}{2005}).

\bibitem[{\citenamefont{Vladkov and Barrat}(2006)}]{vladkov2006}
\bibinfo{author}{\bibfnamefont{M.}~\bibnamefont{Vladkov}} \bibnamefont{and}
  \bibinfo{author}{\bibfnamefont{J.-L.} \bibnamefont{Barrat}},
  \bibinfo{journal}{Macromol. Theor. Simul.} \textbf{\bibinfo{volume}{15}},
  \bibinfo{pages}{252} (\bibinfo{year}{2006}).

\bibitem[{\citenamefont{Tseng et~al.}(2010)\citenamefont{Tseng, Wu, and
  Chang}}]{tseng2010}
\bibinfo{author}{\bibfnamefont{H.-C.} \bibnamefont{Tseng}},
  \bibinfo{author}{\bibfnamefont{J.-S.} \bibnamefont{Wu}}, \bibnamefont{and}
  \bibinfo{author}{\bibfnamefont{R.-Y.} \bibnamefont{Chang}},
  \bibinfo{journal}{Phys. Chem. Chem. Phys.} \textbf{\bibinfo{volume}{12}},
  \bibinfo{pages}{4051} (\bibinfo{year}{2010}).

\bibitem[{\citenamefont{Wittmer et~al.}(2015)\citenamefont{Wittmer, Xu,
  Benzerara, and Baschnagel}}]{wittmer2015}
\bibinfo{author}{\bibfnamefont{J.}~\bibnamefont{Wittmer}},
  \bibinfo{author}{\bibfnamefont{H.}~\bibnamefont{Xu}},
  \bibinfo{author}{\bibfnamefont{O.}~\bibnamefont{Benzerara}},
  \bibnamefont{and}
  \bibinfo{author}{\bibfnamefont{J.}~\bibnamefont{Baschnagel}},
  \bibinfo{journal}{Molecular Physics} \textbf{\bibinfo{volume}{113}},
  \bibinfo{pages}{2881} (\bibinfo{year}{2015}).

\bibitem[{\citenamefont{Bruckner}(1970)}]{bruckner1970}
\bibinfo{author}{\bibfnamefont{R.}~\bibnamefont{Bruckner}},
  \bibinfo{journal}{J. Non-Cryst. Solids} \textbf{\bibinfo{volume}{5}},
  \bibinfo{pages}{123} (\bibinfo{year}{1970}).

\bibitem[{\citenamefont{Zeller and Pohl}(1971)}]{zeller1971}
\bibinfo{author}{\bibfnamefont{R.~C.} \bibnamefont{Zeller}} \bibnamefont{and}
  \bibinfo{author}{\bibfnamefont{R.~O.} \bibnamefont{Pohl}},
  \bibinfo{journal}{Phys. Rev. B} \textbf{\bibinfo{volume}{4}},
  \bibinfo{pages}{2029} (\bibinfo{year}{1971}).

\bibitem[{\citenamefont{Vollmayr et~al.}(1996)\citenamefont{Vollmayr, Kob, and
  Binder}}]{vollmayr1996}
\bibinfo{author}{\bibfnamefont{K.}~\bibnamefont{Vollmayr}},
  \bibinfo{author}{\bibfnamefont{W.}~\bibnamefont{Kob}}, \bibnamefont{and}
  \bibinfo{author}{\bibfnamefont{K.}~\bibnamefont{Binder}},
  \bibinfo{journal}{Phys. Rev. B} \textbf{\bibinfo{volume}{54}},
  \bibinfo{pages}{15808} (\bibinfo{year}{1996}).

\bibitem[{\citenamefont{Taraskin and Elliott}(1999)}]{taraskin1999}
\bibinfo{author}{\bibfnamefont{S.~N.} \bibnamefont{Taraskin}} \bibnamefont{and}
  \bibinfo{author}{\bibfnamefont{S.~R.} \bibnamefont{Elliott}},
  \bibinfo{journal}{J. Phys-Condens. Mat.} \textbf{\bibinfo{volume}{11}},
  \bibinfo{pages}{A219} (\bibinfo{year}{1999}).

\bibitem[{\citenamefont{Rahmani et~al.}(2003)\citenamefont{Rahmani, Benoit, and
  Benoit}}]{rahmani2003}
\bibinfo{author}{\bibfnamefont{A.}~\bibnamefont{Rahmani}},
  \bibinfo{author}{\bibfnamefont{M.}~\bibnamefont{Benoit}}, \bibnamefont{and}
  \bibinfo{author}{\bibfnamefont{C.}~\bibnamefont{Benoit}},
  \bibinfo{journal}{Phys. Rev. B} \textbf{\bibinfo{volume}{68}},
  \bibinfo{pages}{184202} (\bibinfo{year}{2003}).

\bibitem[{\citenamefont{Koziatek et~al.}(2015)\citenamefont{Koziatek, Barrat,
  and Rodney}}]{koziatek2015}
\bibinfo{author}{\bibfnamefont{P.}~\bibnamefont{Koziatek}},
  \bibinfo{author}{\bibfnamefont{J.-L.} \bibnamefont{Barrat}},
  \bibnamefont{and} \bibinfo{author}{\bibfnamefont{D.}~\bibnamefont{Rodney}},
  \bibinfo{journal}{J. Non-Cryst. Solids} \textbf{\bibinfo{volume}{414}},
  \bibinfo{pages}{7–15} (\bibinfo{year}{2015}).

\bibitem[{\citenamefont{van Beest et~al.}(1990)\citenamefont{van Beest, Kramer,
  and van Santen}}]{beest1990}
\bibinfo{author}{\bibfnamefont{B.~W.~H.} \bibnamefont{van Beest}},
  \bibinfo{author}{\bibfnamefont{G.~J.} \bibnamefont{Kramer}},
  \bibnamefont{and} \bibinfo{author}{\bibfnamefont{R.~A.} \bibnamefont{van
  Santen}}, \bibinfo{journal}{Phys. Rev. Lett.} \textbf{\bibinfo{volume}{64}},
  \bibinfo{pages}{1955} (\bibinfo{year}{1990}).

\bibitem[{\citenamefont{Carr\'{e} et~al.}(2007)\citenamefont{Carr\'{e},
  Berthier, Horbach, Ispas, and Kob}}]{carre2007}
\bibinfo{author}{\bibfnamefont{A.}~\bibnamefont{Carr\'{e}}},
  \bibinfo{author}{\bibfnamefont{L.}~\bibnamefont{Berthier}},
  \bibinfo{author}{\bibfnamefont{J.}~\bibnamefont{Horbach}},
  \bibinfo{author}{\bibfnamefont{S.}~\bibnamefont{Ispas}}, \bibnamefont{and}
  \bibinfo{author}{\bibfnamefont{W.}~\bibnamefont{Kob}}, \bibinfo{journal}{J.
  Chem. Phys.} \textbf{\bibinfo{volume}{127}}, \bibinfo{pages}{114512}
  (\bibinfo{year}{2007}).

\bibitem[{\citenamefont{Mantisi et~al.}(2012)\citenamefont{Mantisi, Tanguy,
  Kermouche, and Barthel}}]{mantisi2012}
\bibinfo{author}{\bibfnamefont{B.}~\bibnamefont{Mantisi}},
  \bibinfo{author}{\bibfnamefont{A.}~\bibnamefont{Tanguy}},
  \bibinfo{author}{\bibfnamefont{G.}~\bibnamefont{Kermouche}},
  \bibnamefont{and} \bibinfo{author}{\bibfnamefont{E.}~\bibnamefont{Barthel}},
  \bibinfo{journal}{Eur. Phys. J. B} \textbf{\bibinfo{volume}{85}},
  \bibinfo{pages}{304} (\bibinfo{year}{2012}).

\bibitem[{\citenamefont{Taraskin and
  Elliott}(1997{\natexlab{b}})}]{taraskin1997b}
\bibinfo{author}{\bibfnamefont{S.~N.} \bibnamefont{Taraskin}} \bibnamefont{and}
  \bibinfo{author}{\bibfnamefont{S.~R.} \bibnamefont{Elliott}},
  \bibinfo{journal}{Phys. Rev. B} \textbf{\bibinfo{volume}{56}},
  \bibinfo{pages}{8605} (\bibinfo{year}{1997}{\natexlab{b}}).

\bibitem[{\citenamefont{Giustino and Pasquarello}(2006)}]{giustino2006}
\bibinfo{author}{\bibfnamefont{F.}~\bibnamefont{Giustino}} \bibnamefont{and}
  \bibinfo{author}{\bibfnamefont{A.}~\bibnamefont{Pasquarello}},
  \bibinfo{journal}{Phys. Rev. Lett.} \textbf{\bibinfo{volume}{96}},
  \bibinfo{pages}{216403} (\bibinfo{year}{2006}).

\bibitem[{\citenamefont{Shcheblanov et~al.}(2015)\citenamefont{Shcheblanov,
  Mantisi, Umari, and Tanguy}}]{shcheblanov2015}
\bibinfo{author}{\bibfnamefont{N.~S.} \bibnamefont{Shcheblanov}},
  \bibinfo{author}{\bibfnamefont{B.}~\bibnamefont{Mantisi}},
  \bibinfo{author}{\bibfnamefont{P.}~\bibnamefont{Umari}}, \bibnamefont{and}
  \bibinfo{author}{\bibfnamefont{A.}~\bibnamefont{Tanguy}},
  \bibinfo{journal}{J. Non-Cryst. Solids} \textbf{\bibinfo{volume}{428}},
  \bibinfo{pages}{6} (\bibinfo{year}{2015}).

\bibitem[{\citenamefont{Allen and Tildesley}(1987)}]{allen1987}
\bibinfo{author}{\bibfnamefont{M.~P.} \bibnamefont{Allen}} \bibnamefont{and}
  \bibinfo{author}{\bibfnamefont{D.~J.} \bibnamefont{Tildesley}},
  \emph{\bibinfo{title}{Computer simulation of liquids}}
  (\bibinfo{publisher}{Clarendon Press}, \bibinfo{year}{1987}).

\bibitem[{\citenamefont{Lutsko}(1989)}]{lutsko1989}
\bibinfo{author}{\bibfnamefont{J.~F.} \bibnamefont{Lutsko}},
  \bibinfo{journal}{J. Appl. Phys.} \textbf{\bibinfo{volume}{65}},
  \bibinfo{pages}{2991} (\bibinfo{year}{1989}).

\bibitem[{\citenamefont{Lema\^{i}tre and Maloney}(2006)}]{lemaitre2006}
\bibinfo{author}{\bibfnamefont{A.}~\bibnamefont{Lema\^{i}tre}}
  \bibnamefont{and} \bibinfo{author}{\bibfnamefont{C.}~\bibnamefont{Maloney}},
  \bibinfo{journal}{J. Stat. Phys.} \textbf{\bibinfo{volume}{123}},
  \bibinfo{pages}{415} (\bibinfo{year}{2006}).

\bibitem[{\citenamefont{Zaccone et~al.}(2011)\citenamefont{Zaccone, Blundell,
  and Terentjev}}]{zaccone2011}
\bibinfo{author}{\bibfnamefont{A.}~\bibnamefont{Zaccone}},
  \bibinfo{author}{\bibfnamefont{J.~R.} \bibnamefont{Blundell}},
  \bibnamefont{and} \bibinfo{author}{\bibfnamefont{E.~M.}
  \bibnamefont{Terentjev}}, \bibinfo{journal}{Phys. Rev. B}
  \textbf{\bibinfo{volume}{84}}, \bibinfo{pages}{174119}
  (\bibinfo{year}{2011}).

\bibitem[{\citenamefont{Taraskin and Elliott}(2000)}]{taraskin2000}
\bibinfo{author}{\bibfnamefont{S.~N.} \bibnamefont{Taraskin}} \bibnamefont{and}
  \bibinfo{author}{\bibfnamefont{S.~R.} \bibnamefont{Elliott}},
  \bibinfo{journal}{Phys. Rev. B} \textbf{\bibinfo{volume}{61}},
  \bibinfo{pages}{12031} (\bibinfo{year}{2000}).

\bibitem[{\citenamefont{Christie et~al.}(2007)\citenamefont{Christie, Taraskin,
  and Elliott}}]{christie2007}
\bibinfo{author}{\bibfnamefont{J.}~\bibnamefont{Christie}},
  \bibinfo{author}{\bibfnamefont{S.}~\bibnamefont{Taraskin}}, \bibnamefont{and}
  \bibinfo{author}{\bibfnamefont{S.}~\bibnamefont{Elliott}},
  \bibinfo{journal}{J. Non-Cryst. Solids} \textbf{\bibinfo{volume}{353}},
  \bibinfo{pages}{2272} (\bibinfo{year}{2007}).

\bibitem[{\citenamefont{Damart et~al.}(2015)\citenamefont{Damart, Giordano, and
  Tanguy}}]{damart2015}
\bibinfo{author}{\bibfnamefont{T.}~\bibnamefont{Damart}},
  \bibinfo{author}{\bibfnamefont{V.~M.} \bibnamefont{Giordano}},
  \bibnamefont{and} \bibinfo{author}{\bibfnamefont{A.}~\bibnamefont{Tanguy}},
  \bibinfo{journal}{Phys. Rev. B} \textbf{\bibinfo{volume}{92}},
  \bibinfo{pages}{094201} (\bibinfo{year}{2015}).

\bibitem[{\citenamefont{DeGiuli
  et~al.}(2014{\natexlab{b}})\citenamefont{DeGiuli, Lerner, Brito, and
  Wyart}}]{degiuli2014b}
\bibinfo{author}{\bibfnamefont{E.}~\bibnamefont{DeGiuli}},
  \bibinfo{author}{\bibfnamefont{E.}~\bibnamefont{Lerner}},
  \bibinfo{author}{\bibfnamefont{C.}~\bibnamefont{Brito}}, \bibnamefont{and}
  \bibinfo{author}{\bibfnamefont{M.}~\bibnamefont{Wyart}},
  \bibinfo{journal}{Proc. Nat. Acad. Sci.} \textbf{\bibinfo{volume}{111}},
  \bibinfo{pages}{17054} (\bibinfo{year}{2014}{\natexlab{b}}).

\end{thebibliography}

\end{document}